\documentclass[journal,twoside,web]{ieeecolor}
\usepackage{generic}
\usepackage{amsmath,amssymb,amsfonts}
\usepackage{algorithmic}
\usepackage{graphicx}
\usepackage{algorithm,algorithmic}
\usepackage{hyperref}
\hypersetup{hidelinks=true}
\usepackage{textcomp}
\usepackage{epsfig}
\usepackage{mathtools}
\makeatletter
\let\IEEEcls@makecaption\@makecaption
\makeatother
\usepackage{subcaption}
\makeatletter
\let\@makecaption\IEEEcls@makecaption
\makeatother
\usepackage{bm}
\usepackage{cite}
\usepackage{etoolbox}
\patchcmd{\thebibliography}{\section*{\color{black}}}{}{}{}

\newif\ifDRAFT
\DRAFTfalse
\ifDRAFT
  \usepackage{soul}
  \usepackage[colorinlistoftodos]{todonotes}
  \usepackage[dvipsnames]{xcolor}
\fi

\def\RR{\mathbb{R}}
\def\SS{\mathbb{S}}
\def\BB{\mathbb{B}}
\def\setA{{\mathcal{A}}}
\def\calA{\mathcal{A}}
\def\calB{\mathcal{B}}
\def\setC{{\mathcal{C}}}
\def\setD{{\mathcal{D}}}
\def\calK{\mathcal{K}}
\def\calL{\mathcal{L}}

\def\tt{^{\mathsf{\scriptscriptstyle T}}}
\def\x{{\rm x}}
\def\vaa{{\rm a}}
\def\sca{{\alpha}}
\def\z{{\rm z}}
\def\e{{\rm e}}

\def\g{{\rm g}}
\def\v{{\rm v}}
\def\w{{\rm w}}
\def\str{^{\star}}
\def\gain{{k}}

\DeclareMathOperator{\spanof}{span}

\DeclareMathOperator{\dom}{dom}
\DeclareMathOperator{\epi}{epi}

\newcommand{\bbm}[1]{\begin{bmatrix} #1 \end{bmatrix}}
\newcommand{\bsm}[1]{\begin{bsmallmatrix} #1 \end{bsmallmatrix}}

\newcommand{\matl}[1]{\begin{matrix*}[l] #1 \end{matrix*}}

\newcommand\norm[1]{\left\lVert#1\right\rVert}

\newcommand{\dintt}[4]{\int_{#1}^{#2}#3\,\mathrm{d}#4}

\newcommand{\pp}[1]{\left({#1}\right)}

\newcommand{\ddt}[1]{\frac{\mathrm{d} {#1}}{\mathrm{d} t}}

\newcommand{\todopink}[1]{}
\newcommand{\todogreen}[1]{}

\ifDRAFT    
    \renewcommand{\todopink}[1]{\todo[inline,color=pink]{#1}}
    \renewcommand{\todogreen}[1]{\todo[inline,color=SpringGreen]{#1}}
    
\fi

\newtheorem{theorem}{Theorem}
\newtheorem{lemma}{Lemma}
\newtheorem{proposition}{Proposition}
\newtheorem{corollary}{Corollary}
\newtheorem{definition}{Definition}
\newtheorem{remark}{Remark}
\newtheorem{assumption}{Assumption}

\definecolor{myblue}{RGB}{0,144,178}
\definecolor{myyellow}{RGB}{230,159,0}
\definecolor{mygreen}{RGB}{0,158,115}
\definecolor{myred}{RGB}{213,94,0}
\definecolor{mymaroon}{RGB}{140,29,64}
\definecolor{gravitygreen}{RGB}{19,137,31}
\definecolor{velocityblue}{RGB}{28,93,207}
\definecolor{thrustred}{RGB}{179,28,53}
\definecolor{myhighlight}{RGB}{255,230,150}
\definecolor{mypurple}{RGB}{191,0,64}

\ifDRAFT    
    \sethlcolor{myhighlight}

\fi

\def\BibTeX{{\rm B\kern-.05em{\sc i\kern-.025em b}\kern-.08em
    T\kern-.1667em\lower.7ex\hbox{E}\kern-.125emX}}
\begin{document}
\title{Practical Complete Tracking Through%
\\Gimbaled Unidirectional Actuation%
\\with Application to Multirotor UAVs%
}
\author{Ian J. Willebeek-LeMair, \IEEEmembership{Student Member, IEEE}, Craig A. Woolsey, \IEEEmembership{Senior Member, IEEE}
\thanks{A preliminary version of this work will appear in the 2026 Conference on Decision and Control and a preprint is available~\cite{willebeek_lemair_practical_2026}. (\textit{Corresponding author: Ian Willebeek-LeMair.}) This work was supported by The Boeing Company. }
\thanks{Ian J. Willebeek-LeMair is a Ph.D. candidate with the Kevin T. Crofton Department of Aerospace and Ocean Engineering at Virginia Tech, Blacksburg, VA 24061 USA (e-mail: ianwl@vt.edu). }
\thanks{Craig A. Woolsey is a Professor with the Kevin T. Crofton Department of Aerospace and Ocean Engineering at Virginia Tech, Blacksburg, VA 24061 USA (e-mail: cwoolsey@vt.edu).}\vspace{-1cm}
}

\maketitle

\begin{abstract}
This paper addresses the problem of practical tracking control for robotic vehicles with gimbaled unidirectional actuators, with special application to multirotor unmanned aerial vehicles.  A widely recognized obstacle to tracking control for multirotors is the free-fall singularity --- when a multirotor is commanded to free fall, the vehicle's target attitude becomes undefined.  This singularity is a pathology pertinent to all robotic vehicles with gimbaled unidirectional actuators.  This paper considers the general theoretical problem for vehicles with gimbaled actuators.  Starting from a baseline robust controller that assumes unconstrained inputs, we redesign the control law to be compatible with the gimbaled actuator.  This is accomplished by driving the output of the gimbaled actuator to a ball centered at the target input value.  The baseline controller's tracking guarantees are recovered in a practical sense.  We apply our results to synthesize a global practical complete position tracking controller for multirotor unmanned aerial vehicles. The theory is demonstrated in simulation.  
\end{abstract}

\begin{IEEEkeywords}
Aerospace control, 
Autonomous vehicles, 
Control theory,
Lyapunov methods,
Nonlinear control systems,
Quadrotors,
Set stability,
Underactuated vehicles.
\end{IEEEkeywords}

\section{Introduction} \label{sec:Introduction}

\begin{figure}[t]
    \centering
    \begin{minipage}{0.4865\textwidth}
        \centering
        \includegraphics[trim={0.0cm 0cm 0.0cm -0.4cm},clip,width=0.99\textwidth]{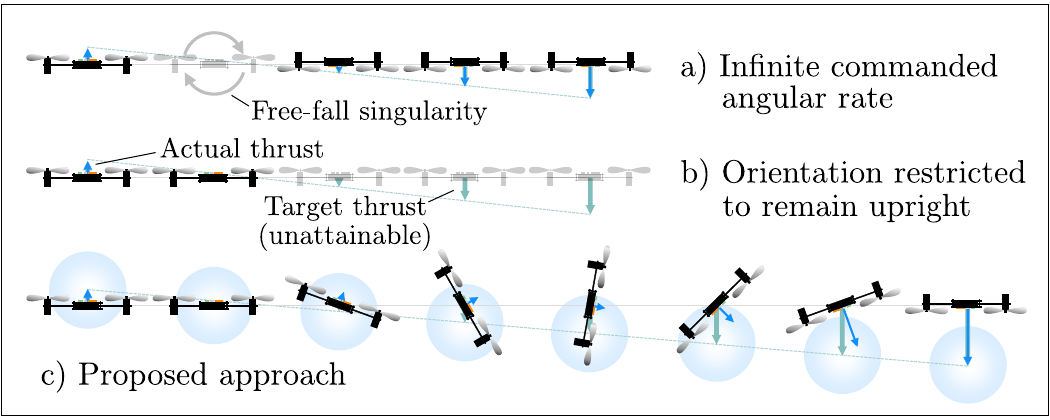}
        \caption{Three multirotor command tracking scenarios.}
        \label{fig:state_of_the_art}
    \end{minipage} 
    \vspace{-0.625cm}
\end{figure}

\IEEEPARstart{T}{he} theoretical problem addressed in this paper is motivated by the applied problem of position and velocity tracking for multirotor unmanned aerial vehicles (UAVs). State-of-the-art nonlinear controllers for multirotors are often designed using backstepping-based approaches that leverage the geometric relationship between a multirotor's orientation and the direction of its thrust vector to synthesize a target attitude from a desired thrust history~\cite{
minh-duc_hua_control_2009,
lee_geometric_2010,
lee_control_2011,
lee_geometric_2011,
mellinger_minimum_2011,
lee_nonlinear_2013,
gamagedara_geometric_2019,
flores_robust_2025,
willebeek_lemair_input_state_2025,
willebeek_lemair_robust_2025,
casau_robust_2015,
naldi_robust_2017,
invernizzi_trajectory_2018,
casau_robust_2019,
invernizzi_dynamic_2020,
martins_global_2021,
martins_robust_2023,
martins_global_2024,
leomanni_almost_2024}.  This target attitude becomes undefined whenever the desired thrust goes through zero, i.e. when the desired motion is free fall, because the vehicle can free fall in any orientation; see Fig.~\ref{fig:state_of_the_art}.  The free-fall singularity obstructs the application of advanced control techniques, such as robust and adaptive control, to all robotic vehicles that employ pivoted or gimbaled unidirectional actuators.  While they are not explored in this work, the results of this paper enable such applications.   

\begin{table}[b!]
    \centering
    \vspace{-0.525cm}
    \caption{Approaches taken to address the free-fall singularity.  }
    \label{tab:free_fall_table}
    \renewcommand{\arraystretch}{1.2}
    \begin{tabular}{lllll}
        \hline
        \hline
        \hspace{-0.2cm} \textbf{Paper(s)} & \hspace{-0.2cm} \textbf{FF Technique} & \hspace{-0.2cm} \textbf{Singular?} & \hspace{-0.2cm} \textbf{Guarantee} & \hspace{-0.2cm} \textbf{Jump?} \\
        \hline
        \cite{minh-duc_hua_control_2009} & \textbf{[A]}~/~\textbf{[R]}+\textbf{[G]} & Yes~/~No & None~/~UUB & Yes \\ \cite{lee_geometric_2010,lee_control_2011,lee_geometric_2011,mellinger_minimum_2011,lee_nonlinear_2013,gamagedara_geometric_2019,flores_robust_2025,willebeek_lemair_input_state_2025,willebeek_lemair_robust_2025} & \textbf{[A]} & Yes & None & Some \\  \cite{casau_robust_2015,naldi_robust_2017,invernizzi_trajectory_2018,casau_robust_2019,invernizzi_dynamic_2020,martins_global_2021,martins_robust_2023,martins_global_2024,leomanni_almost_2024} & \textbf{[R]} & Yes & None & Some \\
        \hline
        This work,~\cite{willebeek_lemair_practical_2026}\hspace{-0.2cm} & \textbf{[S]}+\textbf{[G]} & No & PCT & No \\
        \hline 
        \hline
    \end{tabular}
\end{table}

Recent multirotor control literature addresses the free-fall singularity by either assuming it is never encountered~\cite{minh-duc_hua_control_2009,
lee_geometric_2010,
lee_control_2011,
lee_geometric_2011,
mellinger_minimum_2011,
lee_nonlinear_2013,
gamagedara_geometric_2019,
flores_robust_2025,
willebeek_lemair_input_state_2025,
willebeek_lemair_robust_2025} or excluding any desired position history that includes free fall~\cite{minh-duc_hua_control_2009,
casau_robust_2015,
naldi_robust_2017,
invernizzi_trajectory_2018,
casau_robust_2019,
invernizzi_dynamic_2020,
martins_global_2021,
martins_robust_2023,
martins_global_2024,
leomanni_almost_2024}, typically barring vertical accelerations faster than gravity; see Table~\ref{tab:free_fall_table}. These assumptions and exclusions limit theoretical results.   For example, the assumption that the free-fall singularity is never encountered complicates input-to-state stability (ISS) based robustness analysis since a disturbance could push the vehicle through the singularity, causing a jump discontinuity in the ISS-Lyapunov function. Similarly, limiting the allowable acceleration to be less than gravity precludes ISS results since disturbances might overpower the controller.  

Gimbaled and pivoted unidirectional actuators are often treated as rate-unlimited inputs, even though actuator rotation takes time. Standard backstepping approaches account for the pivot subsystem in the same manner as is used for multirotors and so suffer the same pathology: a desired actuation that goes through zero can demand a discontinuous jump in pivot direction. Inspired by the multirotor scenario, we continue to refer to this singularity as the \textit{free-fall singularity}.  This paper describes a controller redesign that overcomes the singularity and enables tracking without artificial actuation limits. 

Many robotic vehicles employ or themselves are gimbaled or pivoted unidirectional actuators.  A basic example of a pivoted unidirectional actuator is a pinned thruster mounted to a vehicle --- the thruster can only push opposite the direction of its exhaust and it can rotate around the pin.  A more novel example is a rocket with a main thruster mounted to its body --- the thruster can only push opposite the direction of its exhaust and it can be rotated by reorienting the entire vehicle using maneuvering thrusters.  Additional examples include multirotor UAVs, hovercraft that steer with differential thrust, wheeled ground vehicles constrained to forward motion (e.g., a Dubins car), and ocean vehicles with azimuth thrusters. For multirotors, rockets, and hovercraft, the entire vehicle is the gimbal or pivot and the force exerted by the unidirectional actuator is the collective thrust generated by the body-fixed propulsion system~\cite{lee_geometric_2010,ballaben_towards_2025}.  For a wheeled ground vehicle, the body may again be considered the pivot and the unidirectional actuator is forward speed~\cite{bahati_control_2025}. For azimuthal propulsion ocean vehicles, the rotation of the thruster is the pivot and thrust is the unidirectional actuation.  In general, a system with a gimbaled unidirectional actuator is one where a unidirectional control input enters the system as the coefficient of a unit vector whose orientation and angular velocity are controllable continuous states of the system. Our previous work~\cite{willebeek_lemair_practical_2026} addressed the lowest-dimensional case of interest: that of a two-dimensional, i.e. pivoted, unit vector.  (A one-dimensional unit vector cannot change direction continuously.) There are several ways to generalize the pivoted case to gimbaled actuators.  This paper extends our work to the three-dimensional gimbaled case in a manner that is aligned to aerospace and ocean vehicles.  

In prior work~\cite{willebeek_lemair_practical_2026} on pivoted actuators, we proposed a set-based singularity bypass that overcomes the singularity and enables tracking without artificial actuation limits.  Comparable results from the literature are presented in Table~\ref{tab:free_fall_table}.  The table is organized as follows.  The \textit{free-fall technique} column specifies the approach used to address the free-fall singularity: \textbf{[A]} it is assumed that free fall is never encountered,  \textbf{[R]} the reference class is limited to exclude free fall, \textbf{[G]} gates are placed on singular terms, \textbf{[S]} a set-based objective is used. Papers are also marked with whether a singularity is encountered in commanded free fall (\textit{Singular?}), what guarantees persist in that case (\textit{Guarantee}), and whether the associated Lyapunov function can increase via a jump discontinuity at the moment of a commanded free fall (\textit{Jump?}).  Whenever a paper presents multiple controllers, the entry in the table reflects the paper's most sophisticated position-tracking control law.

The closest comparator to both~\cite{willebeek_lemair_practical_2026} and this work in the literature is~\cite{minh-duc_hua_control_2009}, which receives its own row in Table~\ref{tab:free_fall_table}.  That work introduces a controller modification that prevents division by zero at free fall.  For trajectories that include a commanded free fall, the controller ensures uniform ultimate boundedness.  The Lyapunov function used to develop the free-fall-absent asymptotic stability guarantees in~\cite{minh-duc_hua_control_2009}, however, can increase due to a jump discontinuity along some trajectories that pass through free fall.  This work avoids that complication and provides tracking guarantees that are uniform across all reference trajectories.  To the best of the authors' knowledge,~\cite{willebeek_lemair_practical_2026} was the first paper to provide tracking guarantees for a class of reference trajectories that formally includes those passing through free fall.  This paper extends the results.  

In our analysis, we partition a robotic vehicle system into an outer-loop vehicle subsystem whose state may include position, velocity, or both and an inner-loop gimbal subsystem whose state is the gimbal's attitude and angular velocity.  We assume that the gimbaled actuator can exert arbitrarily large positive actuation, that an input-to-state stabilizing exact complete tracking (ECT) controller is available for the outer-loop system, and that the ECT controller's design regards the gimbaled actuator as an unconstrained input.  The controller is then redesigned to be compatible with the constrained motion of a gimbal and to enable practical complete tracking (PCT) for the outer-loop vehicle system.  Exact versus practical complete tracking is clarified in Definition~\ref{def:ECT_and_PCT}.  

This paper extends the results of~\cite{willebeek_lemair_practical_2026} with four major contributions.  First, we address the case of strictly positive unidirectional inputs.  Second, we propose a unidirectional input feedback law that accounts for whether the thrust direction is properly aligned.  Third, we introduce a switching condition that prevents undesirable unwinding while preserving attractivity results.  Fourth, we extend the previous pivoted results to the three-dimensional gimbaled case.

The paper is organized as follows.  Section~\ref{sec:problem_formulation} details the class of systems under consideration, states the objective, and notes an example of an outer-loop trajectory that systems with gimbaled unidirectional actuation cannot follow.  Section~\ref{sec:controller_design} includes the synthesis of the PCT controller and proofs of its properties.  Section~\ref{sec:UAVs} specializes the results to multirotor UAVs. Section~\ref{sec:Simulation} provides a simulation example.  Section~\ref{sec:Conclusions} states the conclusions.

The following notation is used in the paper.  Let $I$ denote an identity matrix, $0$ a matrix of zeros, and $\e_i$ a unit vector with a one in its $i$\textsuperscript{th} entry. The dimensions of $I$, $0$, and $\e_i$ are implied by context.  Let $\norm{\cdot}$ denote the Euclidean norm.  Let $\BB^n$ denote the unit ball in $\RR^n$ with respect to $\|\cdot\|$. For a vector $x\in\RR^n$ and a set $A\subseteq\RR^n$, let $\|x\|_A=\inf_{\z\in A}\|x-\z\|$.  Let $\bar A$, $\partial A$, and ${\rm int}\,A$ denote the closure, the boundary, and the interior of $A$, respectively.  Let $\bar\RR = \RR\cup\{-\infty,\infty\}$ denote the extended real number line.  For any $f:\RR^n\to\RR^m$, let $f(A)=\{v\in\RR^m|v=f(a),a\in A\}$.  Let $C^r(\mathbb{X},\mathbb{Y})$ and ${\rm PWC}(\mathbb{X},\mathbb{Y})$ denote, respectively, the set of $r$-times continuously differentiable functions and the set of piecewise continuous functions from $\mathbb{X}$ to $\mathbb{Y}$.  Let $\wedge:\RR^3\to \mathfrak{so}(3)$ denote the \textit{hat map} operator that maps vectors in $\RR^3$ to cross-product equivalent matrices.  For a vector $a\in\RR^3$, the hat map is variously written as $\hat a$ or $(a)^\wedge$.  Finally, let $\vee: \mathfrak{so}(3)\to\RR^3$ denote the \textit{vee map} operator that is the inverse of the hat map.

\section{Problem Formulation} \label{sec:problem_formulation}

Consider the dynamical system 
\begin{subequations} \label{eq:olsys}
    \begin{align}
        \dot x(t) &= f(t, x(t)) + \g(t, x(t))\lambda(t)u_1(t) 
        \label{eq:olsys_outer_loop_system}
        \\
        \dot\lambda(t) &= (R(t)\omega(t)) \times \lambda(t) 
        \label{eq:olsys_gimbal_pose_1}
        \\
        \dot\mu(t) &= (R(t)\omega(t)) \times \mu(t) 
        \label{eq:olsys_gimbal_pose_2}
        \\
        \dot\omega(t) &= u_2(t) 
        \label{eq:olsys_gimbal_rate}
    \end{align}
\end{subequations}
where $ x\in C^{0}(\RR_{\geq0},\RR^n)$ is the state of the outer-loop system, $f\in C^{\infty}(\RR_{\geq0}\times\RR^n,\RR^n)$ is the smooth drift vector field, $\g\in C^{\infty}(\RR_{\geq0}\times\RR^n,\RR^{n\times3})$ is the smooth actuation matrix which is assumed to be bounded and uniformly full column rank, $\lambda\in C^1(\RR_{\geq0},\SS^2)$ is the direction of the gimbaled unidirectional actuator, $\mu\in C^1(\RR_{\geq0},\SS^2)$ is a second unit vector used to complete the description of the gimbaled unidirectional actuator's orientation, $R\in C^1(\RR_{\geq0},SO(3))$ is the rotation matrix attitude representation $R(t) = [\mu(t) \,\, \lambda(t)\!\times\!\mu(t) \,\, \lambda(t)]$, $\omega\in C^0(\RR_{\geq0},\RR^3)$ is the angular velocity of the gimbal, $u_1\in {\rm PWC}(\RR_{\geq0},[U_1,\infty))$ is the piecewise continuous unidirectional input to the outer-loop system which is lower-bounded by $U_1\geq0$, and $u_2\in {\rm PWC}(\RR_{\geq0},\RR^3)$ is the piecewise continuous angular acceleration input to the gimbal.  

In this paper,~\eqref{eq:olsys} is interpreted as the interconnection of an affine-in-control outer-loop subsystem~\eqref{eq:olsys_outer_loop_system} with an inner-loop gimbal subsystem~\eqref{eq:olsys_gimbal_pose_1},~\eqref{eq:olsys_gimbal_pose_2}, and~\eqref{eq:olsys_gimbal_rate}.  The effective input to the outer-loop system~\eqref{eq:olsys_outer_loop_system} is 
\begin{equation} \label{eq:defn_of_gimbal_thrust}
    \vaa(t) = \lambda(t)u_1(t) \in \RR^3\setminus {\rm int}\,U_1\BB^3
\end{equation}
which we refer to as the ``actuation'' or ``thrust'' applied by the gimbal.  The control-affine structure of~\eqref{eq:olsys_outer_loop_system} and the requirements that $\g$ is bounded and uniformly full column rank are important; they hold for all the vehicle systems mentioned in Section~\ref{sec:Introduction}.  The requirements on $\g$ ensure that the sensitivity of $\dot x(t)$ to changes in $\vaa(t)$ is always both uniformly bounded and nonzero. Boundedness implies small changes in $\vaa(t)$ yield proportionally small changes in $\dot x(t)$, and being nonzero implies that $\vaa(t)$ can always influence $\dot x(t)$.  A nonnegative, possibly nonzero, lower bound $U_1$ for $u_1$ is included because there are practical obstructions to lowering $u_1$ to zero for some vehicle systems.  The motivation for omitting an upper bound for $u_1$ is addressed in Remark~\ref{remark:no_upper_bound}, following the statement of the control objective.  

An example of the system~\eqref{eq:olsys} used throughout this paper is a multirotor aircraft, as depicted in Fig.~\ref{fig:state_of_the_art}. For the multirotor example, equations~\eqref{eq:olsys_outer_loop_system} represent the translational kinematics and dynamics, which are interpreted as an outer-loop of the overall feedback control system. The unit vectors $\lambda$ and $\mu$ are fixed in the body frame but expressed in the inertial frame. Equations~\eqref{eq:olsys_gimbal_pose_1} and~\eqref{eq:olsys_gimbal_pose_2} describe the attitude kinematics, and~\eqref{eq:olsys_gimbal_rate} describes the (simplified) attitude dynamics. The attitude subsystem is controlled as an inner-loop.  The application of the theory developed in this paper to multirotor aircraft is the focus of Sections~\ref{sec:UAVs} and~\ref{sec:Simulation}.  

To develop control laws for~\eqref{eq:olsys}, the thrust $\vaa(t)\in\RR^3\setminus {\rm int}\,U_1\BB^3$ is often replaced with an unconstrained input $u_\vaa(t)\in\RR^3$, which leads to the fictitious surrogate system
\begin{align}
    \begin{aligned}
        \dot x(t) &= f(t, x(t)) + \g(t, x(t)) u_\vaa(t) + \w(t)
    \end{aligned} \label{eq:surrogate_olsys}
\end{align}
where $u_\vaa:\RR_{\geq0}\to\RR^3$ is the (fictitiously) unconstrained thrust input and $\w\in L^{\infty}(\RR_{\geq0},\RR^n)$ is a bounded disturbance.  The system~\eqref{eq:surrogate_olsys} is a surrogate for the entirety of~\eqref{eq:olsys} (not just~\eqref{eq:olsys_outer_loop_system}) and is fictitious because $\vaa(t)=\lambda(t)u_1(t)$ is replaced with $u_\vaa(t)$ which is neither constrained by the lower bound on $u_1(t)$ nor by the differentiability requirement of $\lambda(t)$. 

Let $X\str$ denote the set of undisturbed ($\w\equiv0$) solutions of~\eqref{eq:surrogate_olsys} arising from initial conditions $x(0)\in\RR^n$ and smooth input histories $u_\vaa \in C^\infty(\RR_{\geq0},\RR^3)$.  Then for any smooth desired trajectory $ x\str\in X\str$,~\eqref{eq:surrogate_olsys} can be rewritten
\begin{equation}
    \dot e(t) = f_{ x\str}(t,e(t)) + \g_{ x\str}(t,e(t))u_\vaa(t) + \w(t)
    \label{eq:error_surrogate_olsys}
\end{equation}
where $e(t)= x(t)- x\str(t)$, $f_{ x\str}(t,e(t))=f(t,e(t)+ x\str(t))-\dot x\str(t)$, and $\g_{ x\str}(t,e(t))=\g(t,e(t)+ x\str(t))$.  

\begin{assumption}
\label{assmp:ISS_control_law}
    Consider the fictitious surrogate system~\eqref{eq:error_surrogate_olsys}. Assume there exists a known smooth control law $\kappa_{ x\str}\in C^{\infty}(\RR_{\geq0}\times\RR^n,\RR^3)$ such that $u_\vaa(t)=\kappa_{ x\str}(t,e(t))$ input-to-state stabilizes~\eqref{eq:error_surrogate_olsys} with respect to $\w$, uniformly in $X\str$.  
\end{assumption}

Hence, there exist $\beta\in\calK\calL$ and $\gamma\in\calK_{\infty}$ such that
\begin{equation}
    \|e(t)\| \leq \beta(\|e(t_0)\|,t-t_0) + \gamma({\sup}_{\tau\geq t_0}\|\w(\tau)\|) \label{eq:outer_clsys_is_pre_ISS}
\end{equation}
for all $ x\str\in X\str$, for each $\w\in L^{\infty}(\RR_{\geq0},\RR^n)$.

The objective in this paper is to synthesize a feedback control law for the inputs $u_1$ and $u_2$ that enables the outer-loop system~\eqref{eq:olsys_outer_loop_system} to track any user-supplied smooth outer-loop trajectory $x\str\in X\str$ in a practical sense.  

\begin{definition} \label{def:ECT_and_PCT}
    Consider two control systems $\Sigma_1$ and $\Sigma_2$.  Let $\mathcal{X}$ denote the state space of $\Sigma_1$.  Suppose that the state space of $\Sigma_2$ is $\mathcal{X}\times \mathcal{Y}$.  Let $\mathcal{X}\str$ denote a set of closed-loop trajectories of $\Sigma_1$.  Suppose there exists a controller for $\Sigma_2$, parameterized by $x\str\in \mathcal{X}\str$, that ensures
    \begin{equation}
        \limsup{}_{t\to\infty}\| x(t)- x\str(t)\| \leq \Delta
        \label{eq:control_objective}
    \end{equation}
    for every $x\str\in \mathcal{X}\str$, where $(x(\cdot),y(\cdot)):\RR_{\geq0}\to \mathcal{X}\times \mathcal{Y}$ is a trajectory of $\Sigma_2$.  If for every $x\str\in \mathcal{X}\str$,~\eqref{eq:control_objective} holds with $\Delta=0$ and $x(0) = x\str(0)$ implies there exists some $y(0)\in \mathcal{Y}$ for which $x \equiv x\str$, then the controller is called an \textit{exact \mbox{$\mathcal{X}\str$-complete} tracking} ($\mathcal{X}\str\!$~ECT) \textit{controller for} $\Sigma_2$.  If~\eqref{eq:control_objective} holds with $\Delta>0$, then it is called a \textit{practical \mbox{$\mathcal{X}\str$-complete} tracking} ($\mathcal{X}\str\!$~PCT) \textit{controller for} $\Sigma_2$.  The terms ECT and PCT are used when the set $\mathcal{X}\str$ and the system $\Sigma_2$ are implied by context.  (The choice of terminology is discussed in Remarks~\ref{remark:complete} and~\ref{remark:practical}.)
\end{definition}

The controller $\kappa_{x\str}$ introduced in Assumption~\ref{assmp:ISS_control_law} is an exact \mbox{$X\str$-complete} tracking control law for the system~\eqref{eq:surrogate_olsys}.  This follows from taking $\Sigma_1$ and $\Sigma_2$ to be~\eqref{eq:surrogate_olsys}, $\mathcal{X}\str$ to be $X\str$, and $\mathcal{Y}$ to be the empty set.  

The control objective is to redesign the ISS ECT controller $\kappa_{ x\str}$ for~\eqref{eq:surrogate_olsys} into a practical \mbox{$X\str$-complete} tracking controller for~\eqref{eq:olsys}.  Relaxing the control objective from ECT to PCT enables a set-based singularity bypass that overcomes the free-fall singularity and delivers complete tracking over $X\str$.  

The motivation to synthesize a PCT controller is that ECT controllers for vehicles with gimbaled unidirectional actuation do not exist.  This fact follows from the existence of $x\str\in X\str$ that are not outer-loop trajectories of~\eqref{eq:olsys}, as formalized and discussed in the following theorem and remark.  

\begin{theorem} \label{thrm:no_ECT}
    There exist no ECT controllers for~\eqref{eq:olsys}.
\end{theorem}

\begin{proof}
    ECT requires the implication $x(0)= x\str(0) \Rightarrow x \equiv x\str$ for any $x\str\in X\str$.  If there exists a trajectory $x\str$ of~\eqref{eq:surrogate_olsys} for which this implication fails, then ECT is impossible.  To prove the theorem, we provide such a trajectory of~\eqref{eq:surrogate_olsys} that is not an outer-loop trajectory of~\eqref{eq:olsys}. 
    
    Denote by $x\str$ a trajectory of~\eqref{eq:surrogate_olsys} that results from an arbitrary initial condition $x\str(0)\in\RR^n$, the input history $u_\vaa(t)=(t-1)\e_3$, and $\w\equiv0$.  The input $u_\vaa$ is smooth, so $x\str\in X\str$. Consider a trajectory $x$ of~\eqref{eq:olsys} with $x(0)=x\str(0)$.  Since $\g$ is uniformly full column rank, $x\equiv x\str$ only if $\lambda(t)u_1(t) = u_\vaa(t)$ for all $t\in\RR_{\geq0}$.  Recalling that $u_1(t)\geq U_1 \geq0$, the gimbal orientation consistent with $u_\vaa(t)$ is $\lambda(t)=-\e_3$ for $t<1$ and $\lambda(t)=\e_3$ for $t>1$.  If $U_1>0$ then $x\not\equiv x\str$ since $u_1(t)\lambda(t) \neq u_\vaa(t)$ whenever $\|u_\vaa(t)\|<U_1$.  If $U_1=0$ then $x \equiv x\str$ requires a jump discontinuity in $\lambda(t)$ at $t=1$, which contradicts $\lambda\in C^1$.  Therefore there exist no input histories $u_1,u_2$ for which $x \equiv x\str$, and ECT is impossible.
\end{proof}

\begin{remark} \label{remark:complete}
    The use of ``complete'' in Definition~\ref{def:ECT_and_PCT} is motivated by the following observation.  Let $X$ denote the set of $x$-histories of~\eqref{eq:olsys} arising from smooth input histories.  A corollary to Theorem~\ref{thrm:no_ECT} is that $X$ is a proper subset of $X\str$.  Compared to $X\str$, the set $X$ has \textit{holes} such as the example output history provided in the proof of Theorem~\ref{thrm:no_ECT}.  ECT and PCT address reference histories drawn from the \textit{complete} set $X\str$ rather than its strict subset $X$. 
\end{remark}

\begin{remark} \label{remark:practical}
    The use of the term  ``practical'' in Definition~\ref{def:ECT_and_PCT} is consistent with \textit{input-to-state practical stability} (ISpS), which indicates a nonzero ultimate bound~\cite{lin_various_1995}.
\end{remark}

\begin{remark} \label{remark:practical_example}
    An example helps clarify what it means for a dynamically feasible trajectory of~\eqref{eq:surrogate_olsys} to not be an outer-loop trajectory of~\eqref{eq:olsys}.  When~\eqref{eq:olsys} represents a multirotor UAV, the surrogate system~\eqref{eq:surrogate_olsys} simplifies to a double integrator whose state represents position and velocity and whose input $u_\vaa$ represents the acceleration resulting from the UAV's thrust.  The set $X\str$ of dynamically feasible trajectories of~\eqref{eq:surrogate_olsys} includes ballistic arcs and dives --- histories along which $u_\vaa$ either remains at zero for a finite time or instantaneously crosses through zero, respectively.  For $U_1=0$: thrust can go to zero, so the ballistic arcs of $X\str$ may be outer-loop trajectories (position and velocity output histories) of the multirotor system~\eqref{eq:olsys}.  For $U_1>0$: thrust is nonzero, so the ballistic arcs of $X\str$ are dynamically feasible trajectories of~\eqref{eq:surrogate_olsys} but not outer-loop trajectories of~\eqref{eq:olsys}.  The counterexample used to prove Theorem~\ref{thrm:no_ECT} is a dive and is illustrated in Fig.~\ref{fig:state_of_the_art}.  Regardless of $U_1$, this dive is not an outer-loop trajectory of the multirotor since it requires either $u_1(t)<U_1$ (if $U_1>0$) or a jump in the $\lambda(t)$ component of attitude (if $U_1=0$).  
\end{remark}

\begin{remark} \label{remark:no_upper_bound}
    In this paper, there is no upper bound on $u_1(t)$ for three reasons.  First, an upper bound on $u_1(t)$ would prohibit PCT --- outer-loop histories arising from arbitrarily large $u_\vaa(t)$ in~\eqref{eq:surrogate_olsys} could not be reproduced.  Second, many trajectories of practical importance do not require systems to operate at the upper limits of their actuators, but do require them to operate at or around zero.  The results of this paper are immediately applicable to this class of outer-loop trajectories.  Third, we are interested in developing ISpS results in future work, which will require unbounded control efforts~\cite{lin_various_1995}.  
\end{remark}

\section{Controller Design} \label{sec:controller_design}

This section details the design of the PCT control law for~\eqref{eq:olsys}.  The inner-loop, outer-loop design process mirrors that of~\cite{willebeek_lemair_practical_2026}: supplied with a desired outer-loop trajectory $ x\str(t)$, the outer loop provides the inner loop with a desired thrust history $\vaa\str(t)=\kappa_{ x\str}(t,e(t))$.  Through backstepping, the inner loop then drives $\vaa(t)=\lambda(t)u_1(t)$ to converge to the ball 
\begin{equation}
    \setA(t)=\vaa\str(t)+r \bar\BB^3,
\end{equation}
where $r>U_1$ is a tolerance for error in the commanded thrust vector. (Near the free-fall singularity, where commanded thrust is zero, this tolerance must allow for the lower bound on $u_1(t)$.) The difference $\vaa(t)-\vaa\str(t)$ is then interpreted as a disturbance to the outer-loop system and practical stabilization results follow.  The inner-loop design is described in Section~\ref{subsec:inner-loop} and the long-term behavior of the composite system is discussed in Section~\ref{subsec:long_term}.

The set-based objective $\vaa(t)\to\setA(t)$ is the core of the control design.  This objective addresses the fact that the thrust command $\vaa\str(t)=0$ is unique for a gimbaled actuator.  Whenever $\vaa\str(t)$ is \textit{nonzero}, there is one particular direction
\begin{equation}
    \lambda\str(t) = \frac{\vaa\str(t)}{\|\vaa\str(t)\|}
\end{equation}
from which the gimbaled actuator can supply the target thrust.  For $\vaa\str(t)=0$, in contrast, there is no unique target direction $\lambda\str(t)$ --- the target $\vaa\str(t)=0$ can be attained with $\lambda(t)$ pointed in \textit{any} direction.  The geometric fact that there is a \textit{set} of admissible directions $\lambda(t)$ when $\vaa\str(t)=0$ motivates the set-based control objective $\vaa(t)\to\setA(t)$. Each $\setA(t)$ implicitly defines a set of admissible thrust directions 
\begin{equation}
    \label{eq:defn_of_Lambda}
    \Lambda(t) = \{\lambda\in\SS^2 \,|\, \text{$u_1 \lambda \in \setA(t)$ for some $u_1\in[U_1,\infty)$}\}.  
\end{equation}
Stated differently, the condition $\lambda(t)\in\Lambda(t)$ guarantees $u_1(t)$ can be selected to ensure
\begin{equation}
    \vaa(t)=u_1(t)\lambda(t)\in\setA(t).
    \label{eq:geometry_1}
\end{equation}
Unlike the single thrust direction $\lambda\str(t)$, the set $\Lambda(t)$ is defined for all values of $\vaa\str(t)$.  In particular, $\Lambda(t)=\SS^2$ includes all possible directions $\lambda(t)$ when $\vaa\str(t)=0$ so that the ball $\setA(t)$ is centered at the origin.  When $\|\vaa\str(t)\|$ is large relative to $r$, $\Lambda(t)$ shrinks to a small neighborhood of $\lambda\str(t)$, recovering the performance of a controller designed to drive $\lambda(t)\to\lambda\str(t)$.  At the expense of inexact tracking, this set-based approach enables the controller to become independent of $\lambda\str(t)$ when $\vaa\str(t)$ is in a bounded neighborhood of zero.

\subsection{Inner-loop Design} \label{subsec:inner-loop}

\subsubsection{Specifying a Sufficient Target Attitude Set} \label{subsubsec:target_theta_set}
While the guarantee~\eqref{eq:geometry_1} is most naturally stated in terms of the set $\Lambda(t)$ of unit vectors due to the geometry of the problem, the target attitude set $\Lambda(t)$ is not well suited to backstepping because of the nontrivial condition in its definition~\eqref{eq:defn_of_Lambda}. To facilitate control design, we parameterize $\Lambda(t)$ using the angle $\theta(t)$ between $\lambda\str(t)$ and $\lambda(t)$:
\begin{equation}
    \theta(t) = \arccos(\lambda\str(t)\tt\lambda(t)).
\end{equation}
Let $\Theta(t)$ denote the set of angles 
\begin{equation}
    \Theta(t) = \{\arccos(\lambda\str(t)\tt\lambda) \,|\, \lambda\in\Lambda(t)\}.
\end{equation}
Then $\theta(t)\!\in\!\Theta(t) \Rightarrow \lambda(t)\!\in\!\Lambda(t)$.  Note $\theta(t)$ is undefined when $\vaa\str(t)=0$, so this case must be treated specially.

\begin{figure}[t!]
    \centering
    \begin{minipage}{0.5\textwidth}
        \begin{subfigure}[t]{0.45\textwidth}
            \centering
            \includegraphics[trim={0.0cm 0cm 0.0cm -0.4cm},clip,width=0.95\textwidth]{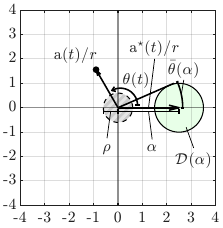}
            \caption{$\sca\in[1+\rho,\infty)$}
            \label{fig:circles_a}
        \end{subfigure}%
        ~
        \begin{subfigure}[t]{0.45\textwidth}
            \centering
            \includegraphics[trim={0.0cm 0cm 0.0cm -0.4cm},clip,width=0.95\textwidth]{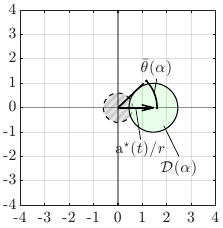}
            \caption{$\sca\in[\sqrt{1+\rho^2},1+\rho)$}
            \label{fig:circles_b}
        \end{subfigure}
    \end{minipage}
    \begin{minipage}{0.5\textwidth}
        \begin{subfigure}[t]{0.45\textwidth}
            \centering
            \includegraphics[width=0.95\textwidth]{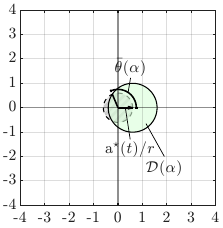}
            \caption{$\sca\in(1-\rho,\sqrt{1+\rho^2})$}
            \label{fig:circles_c}
        \end{subfigure}%
        ~
        \begin{subfigure}[t]{0.45\textwidth}
            \centering
            \includegraphics[width=0.95\textwidth]{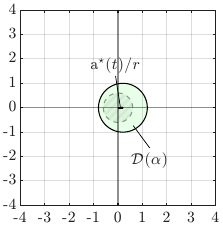}
            \caption{$\sca\in[0, 1-\rho]$}
            \label{fig:circles_d}
        \end{subfigure}
    \end{minipage}
    \caption{Schematic of configurations of $\setD(\sca) = \sca \e_1+\bar\BB^2$ and ${\rm int}\rho\BB^2$.}
    \label{fig:circle_schematic}
\end{figure}

Our aim now is to determine a sufficient condition for $\theta(t) \in \Theta(t)$.  To do this, we introduce a planar surrogate model, normalized by the radius $r$. In this model, $\lambda\str(t)$ defines the $\e_1$-axis and $\lambda(t)$ lies in the $\e_1,\e_2$-plane with a nonnegative $\e_2$ component. The third dimension, which corresponds to $\lambda(t) \times \lambda\str(t)$, is ignored.  This setup is illustrated in Fig.~\ref{fig:circle_schematic} where $\rho = U_1/r\in[0,1)$.  Define
\begin{align}
    \setD(\sca) &= \sca\e_1 + \bar\BB^2, \label{eq:defn_of_D} \\
    L(\phi) &= \{\sigma \, [\cos\phi \,\, \sin\phi]\tt \,| \,\sigma\in\RR\}. \label{eq:defn_of_L}
\end{align}
Without loss of generality, we take $\sca,\phi\geq0$.  With this restriction, $L(\phi)$ intersects $\setD(\sca)$ if and only if
\begin{equation}
    \sca\leq1 \qquad \text{or} \qquad \phi\leq \arcsin(\sca^{-1}). \label{eq:intersection_condition}
\end{equation}
If~\eqref{eq:intersection_condition} holds, then the length along $L(\phi)$ from the origin to the intersection is given by 
\begin{equation}
    l_\pm(\sca,\phi) = \sca \cos\phi \pm \sqrt{1-\sca^2\sin^2\phi}. \label{eq:intersection_length}
\end{equation}
Here, $l_-(\sca,\phi)$ denotes the distance along $L$ to the nearer of two possible intersections with $\setD$, and $l_+(\sca,\phi)$ denotes the distance along $L$ to the farther intersection with $\setD$. For brevity, we refer to these two distances as the ``near-side length'' and the ``far-side length,'' respectively.  The relation~\eqref{eq:intersection_length} is invertible with respect to $\phi$, and its inverse is given by 
\begin{equation}
    \vartheta(\sca,\ell) = \arccos\biggl(\frac{\sca^2+\ell^2-1}{2\sca\ell}\biggr). \label{eq:intersection_angle}
\end{equation}
In this model, the disk $\setD(\sca)$ is a surrogate for $\calA(t)$, the scalar $\sca$ represents the normalized actuation norm $\|\vaa\str(t)\|/r$, and $l_\pm(\sca,\phi)$ correspond to the normalized upper and lower limits on $u_1(t)$ for it to be compatible with $u_1(t)\lambda(t)\in\calA(t)$.  

Toward developing a sufficient condition for $\theta(t) \in \Theta(t)$, consider the set of angles
\begin{equation}
    \Phi(\sca) = \{\phi\in\RR_{\geq0} \,|\, \forall \varphi\in[0,\phi], L(\varphi)\cap(\setD(\sca)\setminus {\rm int}\rho\BB^2)\neq\emptyset\}. \label{eq:defn_of_Phi}
\end{equation}
By construction, $\Phi(\sca)$ is connected, its lower bound is zero, and $\theta(t)\in\Phi(\|\vaa\str(t)\|/r)\Rightarrow\theta(t)\in\Theta(t)$.  To write $\Phi(\sca)$ in interval notation, define
\begin{equation}
    \bar\theta(\sca) = \sup \Phi(\sca). \label{eq:defn_of_thetabar}
\end{equation}
The behavior of $\bar\theta$ can be split into three cases.  First: when $\sca$ is relatively large (i.e., $\sca>\sqrt{1+\rho^2}$), $\setD(\sca)$ is sufficiently far from the origin that $L(\bar\theta(\sca))$ is tangent to $\partial\setD(\sca)$; see Fig.~\ref{fig:circles_a},~\ref{fig:circles_b}.  Second: when $\sca$ is small enough that $l_+(\sca,\arcsin(\sca^{-1}))\leq \rho$ but not so small that $\rho\BB^2 \subset \setD(\sca)$ (i.e., $\sca\in(1-\rho,\sqrt{1+\rho^2}]$), $L(\bar\theta(\sca))$ is the line that passes through the intersection of $\partial\setD(\sca)$ and $\rho\partial\BB^2$ at a distance $\rho$ from the origin; see Fig.~\ref{fig:circles_c}.  Third: when $\sca$ is small enough that $\rho\BB^2 \subset \setD(\sca)$ (i.e., $\sca\in[0,1-\rho]$), the set $\setD(\sca)\setminus {\rm int}\rho\BB^2$ surrounds the origin, so $L(\phi)$ intersects it for any $\phi\geq0$; see Fig.~\ref{fig:circles_d}.  The supremum $\bar\theta(\sca)$ is, therefore, given by
\begin{equation}
    \bar\theta(\sca) = \left\{
    \matl{
    \infty, & \sca\in[0,1-\rho]
    \\
    \vartheta(\sca,\rho), & \sca\in (1-\rho,\sqrt{1+\rho^2})
    \\
    \arcsin(\sca^{-1}), & \sca\geq \sqrt{1+\rho^2}
    }
    \right.. \label{eq:theta_bar}
\end{equation}

\begin{figure}[t!]
    \centering 
    \begin{minipage}{0.48\textwidth}
        \centering
        \includegraphics[trim={0.0cm 0cm 0.0cm 0.0cm},clip,width=0.99\textwidth]{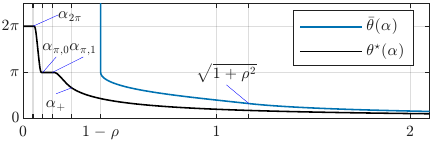}
        \caption{Illustration of $\theta\str(\sca)\leq\bar\theta(\sca)$ for $\rho=0.6$.}
        \label{fig:graphs_of_thetas}
    \end{minipage}\vspace{-0.4cm}
\end{figure}

Recalling that the range of $\theta(t)$ is $[0,\pi]$, the set $\Theta(t)$ can be written as 
\begin{equation}
    \Theta(t) = \Phi(\|\vaa\str(t)\|/r) \cap [0,\pi]= [0,\min\{\bar\theta(\|\vaa\str(t)\|/r),\pi\}]. \label{eq:interval_Theta}
\end{equation}
The upper bound $\min\{\bar\theta(\|\vaa\str(t)\|/r),\pi\}$ does not depend smoothly on $\|\vaa\str(t)\|$, so it is not well suited for backstepping.  The graph of $\bar\theta(\sca)$ in Fig.~\ref{fig:graphs_of_thetas}, however, makes it clear that smooth underestimates exist.

Toward developing a smooth underestimate of $\bar\theta$, note the following property of $l_+(\sca,\phi)$ and $\vartheta(\sca,\ell)$.  For all $\sca\in(0,1]$
\begin{equation}
    \ell\in[l_+(\sca,\pi),l_+(\sca,0)] \Rightarrow \vartheta(\sca,\ell)\in[0,\pi]\subset[0,\bar\theta(\sca)], \label{eq:lengths_to_angles_1}
\end{equation}
and for all $\sca>1-\rho$
\begin{equation}
    \ell\in[l_+(\sca,\bar\theta(\sca)),l_+(\sca,0)] \Rightarrow \vartheta(\sca,\ell)\in[0,\bar\theta(\sca)]. \label{eq:lengths_to_angles_2}
\end{equation}
The implications~\eqref{eq:lengths_to_angles_1} and~\eqref{eq:lengths_to_angles_2} guarantee the existence of an angle corresponding to a far-side intersection.  They both follow from the observation that connected intervals of far-side lengths correspond to connected intervals of intersection angles according to~\eqref{eq:intersection_length} and~\eqref{eq:intersection_angle}.  Implication~\eqref{eq:lengths_to_angles_1} is valid for $\sca\in(0,1]$ because $\vartheta(\sca,\ell)$ is undefined for $\sca=0$ and $(\sca,\pi)$ does not satisfy~\eqref{eq:intersection_condition} for $\sca>1$.  Implication~\eqref{eq:lengths_to_angles_2} is valid for $\sca>1-\rho$ because $l_+(\sca,\bar\theta(\sca))$ is undefined when $\bar\theta(\sca)=\infty$.  These two implications, taken together, can be used to construct a smooth underestimate $\theta\str(\sca)$ of $\bar\theta(\sca)$ from an appropriate smooth far-side length profile $l\str_+(\sca)$.  Toward this construction, note that $l_+(\sca,0) = \sca+1$ for $\sca\geq0$, that $l_+(\sca,\pi) = 1-\sca$ for $\sca\in[0,1]$, and that for $\sca > 1-\rho$, 
\begin{equation}
    l_+(\sca,\bar\theta(\sca)) = \left\{
    \matl{
    \rho, & \sca\in(1-\rho,\sqrt{1+\rho^2}) \\
    \sca\cos\bar\theta(\sca), & \sca\geq \sqrt{1+\rho^2}
    }
    \right.,
\end{equation}
from which it follows that
\begin{equation}
    l_+(\sca,\bar\theta(\sca)) \leq \max\{\rho,\sca\} \qquad \forall\, \sca\geq 1-\rho. \label{eq:length_overestimate}
\end{equation}
To fit the existence criteria~\eqref{eq:lengths_to_angles_1} and~\eqref{eq:lengths_to_angles_2}, therefore, it is sufficient for $l\str_+(\sca)$ to satisfy
\begin{equation}
    l\str_+(\sca) \in \left\{
    \matl{
    [1-\sca,\sca+1], & \sca\in[0,1-\rho] \\
    [\max\{\rho,\sca\},\sca+1], & \sca>1-\rho
    }
    \right.. \label{eq:l_star_existence_condition}
\end{equation}
To this sufficient condition, we choose to add two design targets that enable desirable properties.  First, we choose to design $l\str_+(\sca)$ to satisfy
\begin{equation}
    l\str_+(\sca)=1-\sca \qquad \forall\,\sca\in[0,\sca_{\pi,1}] \label{eq:optional_l_star_condition_1}
\end{equation}
for some $\sca_{\pi,1}\in(0,1-\rho)$.  This choice ensures $\vartheta(\sca,l\str_+(\sca))=\pi$ for all $\sca$ in a finite neighborhood of zero and has two desirable effects. It (i) ensures all gimbal orientations eventually become admissible (i.e., $\theta(t)\leq\vartheta(0,l\str_+(0))=\pi$ holds for any $\lambda(t),\lambda\str(t)\in\SS^2$), and it (ii) ensures that the derivative of $\vartheta(\cdot,l\str_+(\cdot))$ is uniformly zero on a finite neighborhood of zero.  The latter property is exploited in a modification introduced at the end of this section.  Second, we choose to design $l\str_+(\sca)$ to satisfy
\begin{equation}
    l\str_+(\sca)\geq \max\Bigl\{\frac{1+\rho}{2},\sca\Bigr\} \qquad \forall\,\sca\geq0. \label{eq:optional_l_star_condition_2}
\end{equation}
This design choice facilitates the design of the feedback law for $u_1(t)$ in Section~\ref{subsec:u1_law}.  

To ensure that $\theta\str(\sca)$ and $l\str_+(\sca)$ are smooth, the prescription for $l\str_+(\sca)$ involves a two-parameter function $Z(s;s_0,s_1)$, defined in Appendix~\ref{app:zeta}, that smoothly approximates a unit ramp function. For any $s_0 < s_1$, $Z(s;s_0,s_1)=0$ for $s\leq s_0$, then it rises smoothly from zero to $(s_1-s_0)/2$ over $s\in[s_0,s_1]$, and takes the value $Z(s;s_0,s_1)=s-(s_1+s_0)/2$ for $s\geq s_1$.  Define $l\str_+:\RR_{\geq0}\to\RR_{\geq0}$ as 
\begin{equation}
    l\str_+(\sca) = 1-\sca + 2Z(\sca;\sca_{\pi,1},\sca_+),
\end{equation}
with $\sca_{\pi,1}=\tfrac{3}{4}\tfrac{1-\rho}{2}$ and $\sca_+ = \tfrac{5}{4}\tfrac{1-\rho}{2}$.  By construction, $l\str_+(\sca)$ satisfies~\eqref{eq:l_star_existence_condition},~\eqref{eq:optional_l_star_condition_1}, and~\eqref{eq:optional_l_star_condition_2}.  

The intersection angle profile corresponding to $l\str_+$ is
\begin{equation}
    \vartheta(\sca,l\str_+(\sca)) = \arccos\biggl(\frac{\sca^2+l\str_+(\sca)^2-1}{2\sca\,l\str_+(\sca)}\biggr), \label{eq:proto_theta_star}
\end{equation}
where we define $\vartheta(0,l\str_+(0)) = \lim{}_{\sca\to0}\vartheta(\sca,l\str_+(\sca))=\pi$.  This function is smooth and underestimates $\bar\theta$ by construction.  Instead of using $\vartheta(\cdot,l\str_+(\cdot))$ directly as $\theta\str$, however, we prescribe
\begin{equation} \label{eq:defn_of_theta_star}
    \begin{aligned}
        \theta\str(\sca) &= \vartheta(\sca,l\str_+(\sca)) \zeta(\sca;\sca_{2\pi},\sca_{\pi,0}) \\&\qquad {} + 2\pi (1-\zeta(\sca;\sca_{2\pi},\sca_{\pi,0})), 
    \end{aligned}
\end{equation}
with $\sca_{2\pi}=\tfrac{1}{4}\tfrac{1-\rho}{2}$ and $\sca_{\pi,0}=\tfrac{2}{4}\tfrac{1-\rho}{2}$; where $\zeta(s;s_0,s_1)$ is a smooth two-parameter step function related to $Z$ that transitions from zero at $s=s_0$ to one at $s=s_1$ and is described in Appendix~\ref{app:zeta}.  The design of~\eqref{eq:defn_of_theta_star} ensures $\theta\str(\sca)$ takes a value of $2\pi$ in a finite neighborhood of zero.  This modification makes use of~\eqref{eq:optional_l_star_condition_1}: as $\sca$ decreases toward zero, the value of $\theta\str(\sca)$ smoothly converges to $\pi$ at $\sca_{\pi,1}$, holds this value over $\sca\in[\sca_{\pi,0},\sca_{\pi,1}]$, smoothly transitions to $2\pi$ at $\sca_{2\pi}$, and then holds that value over $\sca\in[0,\sca_{2\pi}]$.  The design choice to add the smooth transition to $2\pi$ is used in Section~\ref{subsec:introducing_MRPs} to admit every modified Rodrigues parameter (a double covering of orientations)  into the allowable set whenever $\vaa\str(t)$ is in a small, \emph{finite} neighborhood of zero.  

The main results of this section are the two following lemmas which establish $\theta\str$ is, indeed, an underestimate of $\bar\theta$ and state a sufficient condition for $\theta(t)\in\Theta(t)$.  

\begin{lemma} \label{lemma:theta_star_lemma}
    The function $\theta\str$ is well-defined, positive, and underestimates $\bar\theta(\sca)$ for all $\sca\geq0$.  
\end{lemma}

\begin{proof}
    Consider $\vartheta(\cdot,l\str_+(\cdot))$.  According to the definition of $\vartheta(\cdot,l\str_+(\cdot))$ in~\eqref{eq:l_star_existence_condition}, $l\str_+(\sca)$ satisfies~\eqref{eq:lengths_to_angles_1} and~\eqref{eq:lengths_to_angles_2}.  It follows that $\vartheta(\sca,l\str_+(\sca))$ is well-defined, positive, and underestimates $\bar\theta(\sca)$ for all $\sca\geq0$. According to the definition of $\theta\str$ in~\eqref{eq:defn_of_theta_star}, $\theta\str(\sca)\in(\vartheta(\sca,l\str_+(\sca)),2\pi]$ for $\sca\in[0,\sca_{\pi,0})$ and $\theta\str(\sca)=\vartheta(\sca,l\str_+(\sca))$ for $\sca\geq\sca_{\pi,0}$.  The claim immediately follows.  
\end{proof}

\begin{lemma} \label{lemma:sufficient_for_ball}
    Whenever $\vaa\str(t)$ is nonzero, 
    \begin{equation}
        \theta(t)\in\Theta\str(t)\coloneqq [0,\theta\str(\|\vaa\str(t)\|/r)]
    \end{equation}
    implies there exists $u_1(t)\geq U_1$ for which~\eqref{eq:geometry_1} holds.  
\end{lemma}

\begin{proof}
    Suppose $\vaa\str(t)\neq0$ and $\theta(t)\in\Theta\str(t)$.  If $\theta(t)=0$, then $\max\{U_1,\|\vaa\str(t)\|\}\lambda(t)\in\calA(t)$.  Otherwise if $\theta(t)\neq0$, let $\lambda\str_\perp(t)$ be the unit vector in $\lambda(t),\lambda\str(t)$-plane which is orthogonal to $\lambda\str(t)$ and satisfies $\lambda(t)\tt\lambda\str_\perp(t)>0$. Then $\lambda(t)=\cos\theta(t)\lambda\str(t)+\sin\theta(t)\lambda\str_\perp(t)$ and~\eqref{eq:intersection_condition} ensures the line $\{\lambda(t)u_1 \,|\,u_1\geq0\}$ intersects $\calA(t)$.  Along this line, $r>U_1$ ensures there exists $u_1(t)\geq U_1$ for which~\eqref{eq:geometry_1} holds.  
\end{proof}

\subsubsection{Specifying the Auxiliary Direction} \label{subsec:aux_direction}
The auxiliary direction $\mu(t)$, which completes the gimbal orientation, remains to be addressed.  Unlike $\lambda\str(t)$ for $\lambda(t)$, the outer-loop controller does not provide a desired value for $\mu(t)$.  Instead, we assume a desired pointing vector $\nu:\RR_{\geq0}\to\RR^3$ is provided.  The secondary control objective, after ensuring~\eqref{eq:control_objective}, is to encourage
\begin{equation}
    \mu(t) \to \spanof\{\vaa\str(t),\nu(t)\} \quad \text{s.t.} \quad \mu(t)\tt\nu(t)\geq0
    \label{eq:secondary_control_objective}
\end{equation}
whenever the pair $\{\vaa\str(t),\nu(t)\}$ is linearly independent.  

\begin{remark}
    For many applications, it is useful to be able to influence $\mu(t)$.  When working with multirotor UAVs, for example, $\mu(t)$ often corresponds to the line of sight of a first-person-view camera.  The control objective~\eqref{eq:secondary_control_objective} is selected, in particular, because it is compatible with backstepping.  The objective~\eqref{eq:secondary_control_objective}, rather than the more common $\mu(t)\to\nu(t)/\|\nu(t)\|$, is often sufficient in practice: $\vaa\str(t)$ is upright for a multirotor in hover, so specifying $\nu(t)$ to be any nonzero vector in the plane of the horizon leads to $\mu(t)\to\nu(t)/\|\nu(t)\|$ as $\lambda(t)\to\lambda\str(t)$.  Our approach is distinct from others in the literature because $\nu(t)$ is not necessarily unit norm, $\nu(t)$ is not confined to the plane of the horizon, and the objective~\eqref{eq:secondary_control_objective} disengages if $\nu(t)$ goes to zero causing $\{\vaa\str(t),\nu(t)\}$ to be linearly dependent. 
\end{remark}

\begin{remark}
    Other works assume a target value $\mu\str(t)$ for $\mu(t)$ is directly provided or they reduce the gimbal's attitude description to two angles~\cite{kim_global_2026} (i.e., retain $\lambda(t)$ and omit $\mu(t)$).  These approaches are not employed here because they involve either placing algebraic constraints on $\mu\str(t)$ that would impede the backstepping process later in the design or eliminating the controller's ability to influence $\mu(t)$.
\end{remark}

Toward ensuring~\eqref{eq:secondary_control_objective}, temporarily suppose there exists some differentiable $\mu\str:\RR_{\geq0}\to\SS^2$ whose values are at all times orthogonal to $\lambda\str(t)$.  Several conditions on $\dot\mu\str(t)$ arise from the fact that $\{\mu\str(t),\lambda\str(t)\!\times\!\mu\str(t),\lambda\str(t)\}$ is a smoothly varying orthonormal basis for $\RR^3$ whenever $\lambda\str(t)$ is defined.  Along a time interval where $\vaa\str(t)\neq0$, so that $\lambda\str(t)$ is well-defined,
\begin{align}
    \mu\str(t)\tt \mu\str(t) = 1 \,\, &\Rightarrow \,\, \mu\str(t)\tt \dot\mu\str(t) = 0,
    \\
    \lambda\str(t)\tt \mu\str(t) = 0 \,\, &\Rightarrow \,\, \lambda\str(t)\tt \dot\mu\str(t) = - \mu\str(t)\tt \dot\lambda\str(t).
\end{align}
It follows that whenever $\lambda\str(t)$ is defined,
\begin{equation}
    \dot\mu\str(t) = \Omega_{\mu}(t) \lambda\str(t)\!\times\!\mu\str(t) - \lambda\str(t)\mu\str(t)\tt \dot\lambda\str(t).
    \label{eq:necessary_for_mu_str_dot}
\end{equation}
While~\eqref{eq:necessary_for_mu_str_dot} is derived as a necessary condition, it can be used directly in the control design, alleviating the assumption that a suitable $\mu\str$ exists.  

\begin{lemma}
    Prescribe $\mu\str(t)$ to be a state of the controller which evolves according to~\eqref{eq:necessary_for_mu_str_dot}.  Suppose that $\mu\str(0)$ is selected so that it is a unit vector orthogonal to $\lambda\str(0)$.  Then $\|\mu\str(t)\|=1$ and $\lambda\str(t)\tt \mu\str(t) = 0$ for all $t\in\RR_{\geq0}$.  
\end{lemma}

\begin{proof}
    Both $\|\mu\str(t)\|^2$ and $\lambda\str(t)\tt \mu\str(t)$ are integrals of motion of~\eqref{eq:necessary_for_mu_str_dot} since $\ddt{}\mu\str(t)\tt \mu\str(t)\!=\!2\mu\str(t)\tt \dot\mu\str(t)\!=\!0$ and $\ddt{}\lambda\str(t)\tt \mu\str(t)\!=\!\mu\str(t)\tt\dot\lambda\str(t)\!+\!\lambda\str(t)\tt\dot\mu\str(t)\!=\!0$. Hence, $\mu\str(t)$ remains a unit vector orthogonal to $\lambda\str(t)$ for all time.  
\end{proof}

Beyond the prescription above, two important observations emerge from~\eqref{eq:necessary_for_mu_str_dot}.  First, the free term $\Omega_{\mu}(t)$ can be used to design the evolution of $\mu\str(t)$.  Second, if $\lambda\str(t)$ ever becomes undefined (i.e., if ever $\vaa\str(t)=0$), then $\mu\str(t)$ necessarily becomes undefined as well.  Leveraging the first point, prescribe
\begin{equation}
    \Omega_{\mu}(t) = \gain_{\mu} (\lambda\str(t)\times\mu\str(t))\tt\nu(t).
\end{equation}
This choice of $\Omega_\mu(t)$ drives $\mu\str(t)$ toward the plane shared by $\vaa\str(t)$ and $\nu(t)$ and it tends to stabilize $\mu\str(t)\tt\nu(t)\geq0$.  Hence, $\mu\str(t)$ converges to $\spanof\{\vaa\str(t),\nu(t)\}$, so convergence of $\mu(t)$ to $\mu\str(t)$ will yield the desired behavior~\eqref{eq:secondary_control_objective}.  

The targets $\lambda\str(t)$ and $\mu\str(t)$ define a target rotation matrix,
\begin{equation}
    R\str(t) = \bbm{\mu\str(t) & \lambda\str(t)\times\mu\str(t) & \lambda\str(t)},
\end{equation}
which has the associated angular velocity
\begin{align}
    \omega_{R\str}\!(t) &= (R\str(t)\tt \dot R\str(t))^{\vee} \notag \\
    &= (\e_2\mu\str(t)\tt \!- \e_1(\lambda\str(t)\!\times\!\mu\str(t))\tt)\smash{\frac{\dot\vaa\str(t)}{\|\vaa\str(t)\|}} + \Omega_\mu(t)\e_3 \notag
    \\
    &= R\str(t)\tt\pp{\lambda\str(t)\times\frac{\dot\vaa\str(t)}{\|\vaa\str(t)\|} + \lambda\str(t) \Omega_\mu(t)}. \label{eq:defn_of_omega_R_star} 
\end{align}

\begin{remark}
    The expression on the right hand side of~\eqref{eq:defn_of_omega_R_star} appears in the upcoming backstepping analysis.  When $\vaa\str(t)=0$, $\omega_{R\str}(t)$ is undefined since $\lambda\str(t)$ is undefined.  When $\vaa\str(t)$ approaches zero, $\omega_{R\str}(t)$ typically blows up toward infinity due to the division by $\|\vaa\str(t)\|$.  It is possible for $\vaa\str(t)$ to approach zero without $\omega_{R\str}(t)$ diverging: if $\vaa\str(t)$ and $\dot\vaa\str(t)$ are parallel during the zero crossing, as in every row of Fig.~\ref{fig:state_of_the_art}, then $\omega_{R\str}(t)$ can remain bounded while $\vaa\str(t)$ approaches or leaves zero.  In this circumstance, however, there is a jump discontinuity in $R\str(t)$ when $\vaa\str(t) = 0$.  A Lyapunov-based analysis that relies on differences between $R$ and $R\str$, $\omega$ and $\omega_{R\str}$, or both may appear to guarantee good performance while ignoring mathematically unallowable and physically unrealizable jump discontinuities --- standard Lyapunov arguments require that a candidate Lyapunov function $V(t,x)$ be continuous in time~\cite{khalil_nonlinear_2002}.  The set-based approach used in this paper avoids the discontinuity and the blow-up by ensuring the controller does not depend on $\vaa\str(t)$ whenever $\|\vaa\str(t)\|$ is sufficiently small.  
\end{remark}

Returning to the second takeaway from~\eqref{eq:necessary_for_mu_str_dot}: if $\mu\str(t)$ can become undefined like $\lambda\str(t)$, then it is appropriate to use the same set-based approach for $\mu(t)$ as for $\lambda(t)$.  Toward that end, we define the error rotation matrix 
\begin{equation}
    \tilde R(t) = R\str(t)\tt R(t)
\end{equation}
whose derivative, when it is well-defined, can be written in terms of the angular velocity $\tilde\omega(t) = \omega(t)-\tilde R(t)\tt \omega_{R\str}(t)$.

\subsubsection{Using MRPs for Orientation} \label{subsec:introducing_MRPs}
In the following, the error rotation matrix $\tilde R(t)$ is parameterized by the modified Rodrigues parameter (MRP) vector $\eta(t)$. Define $\eta:\RR_{\geq0}\to\bar\RR^3$ by continuity and the inclusion
\begin{equation}
    \eta(t)\in\left\{E(\tilde R(t)),-\frac{E(\tilde R(t))}{\|E(\tilde R(t))\|^2}\right\},
    \label{eq:defn_of_eta}
\end{equation}
where $E:SO(3)\to\bar\BB^3$ is the mapping
\begin{equation}
    E(R) = \tan(\tfrac{1}{4}\psi(R))\,\mathfrak{a}(R)
\end{equation}
wherein $\mathfrak{a}:SO(3)\to\SS^2$ and $\psi:SO(3)\to[-\pi,\pi)$ map rotation matrices into corresponding Euler axis-angle representations; i.e., $R = e^{\psi(R) \hat{\mathfrak{a}}(R)}$.  The initial condition of $\eta(t)$ is taken to be $\eta(0)=E(\tilde R(0))\in\bar\BB^3$ if $\lambda\str(0)$ and $\mu\str(0)$ are defined.  The case where $\lambda\str(0)$ and $\mu\str(0)$ are not defined is addressed by a switching rule presented in Section~\ref{subsec:switching_condition}.  

The set of MRPs is $\bar\RR^3$ and is a double covering of $SO(3)$.  Each MRP can be mapped to a \textit{shadow MRP} that corresponds to the same rotation.  The function that converts between MRPs and shadow MRPs is ${\rm sh}(\eta) = -\eta/\|\eta\|^2$. The extended reals are used in the definition of MRPs rather than the reals because MRPs that include infinite entries are well-defined and correspond to $R(t)=R\str(t)$ owing to the equivalence of MRPs and their shadows.  With MRP shadows defined, we extend the domain of $E^{-1}$ from $\BB^3$ to $\bar\RR^3$ with $E^{-1}(\eta)=E^{-1}({\rm sh}(\eta))$ for all $\eta\in\bar\RR^3$. 

Using the coordinate $\eta(t)$ to describe the orientation of the gimbal resolves the topological obstruction encountered when $R(t)$ and $R\str(t)$ are $180^{\circ}$ away from one another~\cite{liberzon_switching_2003}.  The coordinate $\eta$ provides a double covering of $SO(3)$. Rotations that are $180^{\circ}$ away from $R(t)=R\str(t)$ are mapped to pairs of antipodal points on the unit sphere.  The topological obstruction is then resolved by selecting the angular velocity to drive whichever MRP is active toward the origin.  Sometimes, it is appropriate to switch the MRP representing the gimbal orientation, as explained in Section~\ref{subsec:switching_condition}.  When it is finite and not switching, $\eta(t)$ evolves continuously according to 
\begin{equation}
    \dot\eta(t) = T(\eta(t))\tilde\omega(t)=T(\eta(t))\pp{\omega(t)-\tilde R(t)\tt \omega_{R\str}(t)},
    \label{eq:defn_of_ddt_eta} 
\end{equation}
where $T:\bar\RR^3\to\bar\RR^{3\times3}$ is defined by
\begin{equation}
    \hspace{-0.15cm}
    T(\eta) = \tfrac{1}{4}\bigl((1-\norm{\eta}^2)I + 2 \hat{\eta} + 2 \eta\eta\tt\bigr), \label{eq:defn_of_T}
\end{equation}
which follows from taking the derivative of~\eqref{eq:defn_of_eta}.  When $\eta(t)\in \bar\RR^3\!\setminus\!\RR^3$, its shadow is zero and $\ddt{}{\rm sh}(\eta(t))=\tfrac{1}{4}\tilde\omega(t)$.  The next lemma defines a desired set of attitudes that is used in the following control design process.  

\begin{lemma} \label{lemma:bound_on_eta}
    Let $n\str(t) = \tan(\tfrac{1}{4}\theta\str(\|\vaa\str(t)\|/r))$.  Suppose
    \begin{equation}
        \eta(t)\in N\str(t)\coloneqq n\str(t)\bar\BB^3.
        \label{eq:bound_on_eta}
    \end{equation}
    Then both $\theta(t)$ and $\arccos(\mu(t)\tt\mu\str(t))$ are less than or equal to $\theta\str(\|\vaa\str(t)\|/r)$.  
\end{lemma}

\begin{proof}
    The rotation $\eta(t)$ is at an angle $4\arctan(\|\eta(t)\|)$ away from the identity rotation.  This implies that $\tilde R(t) = \exp(4\arctan(\|\eta(t)\|)(\eta(t)/\|\eta(t)\|)^\wedge) = R\str(t)\tt R(t)$ where we have used an Euler axis-angle representation. The angle between corresponding columns of $R(t)$ and $R\str(t)$ is therefore upper bounded by the Euler angle $4\arctan(\|\eta(t)\|)$.  Hence, 
    \begin{equation}
        \max\{\theta(t),\arccos(\mu(t)\tt\mu\str(t))\} \leq 4\arctan(\|\eta(t)\|).
        \label{eq:bound_on_both_angles}
    \end{equation}
    In addition, the inclusion assumed in~\eqref{eq:bound_on_eta} implies that
    \begin{equation}
        \label{eq:inclusion_to_bound}
        4\arctan(\|\eta(t)\|) \leq 4\arctan(n\str(t)) = \theta\str(\|\vaa\str(t)\|/r)
    \end{equation}
    where we have used the definition of $n\str(t)$ given in Lemma~\ref{lemma:bound_on_eta}.  Pairing~\eqref{eq:bound_on_both_angles} with~\eqref{eq:inclusion_to_bound} yields the claimed result.
\end{proof}

The meaning of Lemma~\ref{lemma:bound_on_eta} is similar to that of Lemma~\ref{lemma:theta_star_lemma}: ensuring $\|\eta(t)\| \leq n\str(t)$ guarantees the existence of some $u_1(t)\geq U_1$ for which the objective~\eqref{eq:geometry_1} holds. Lemmas~\ref{lemma:theta_star_lemma} and~\ref{lemma:bound_on_eta} are different in one important respect: the guarantee resulting from Lemma~\ref{lemma:bound_on_eta} concerns the complete attitude of the vehicle so $\|\eta(t)\| \leq n\str(t)$ also implies $\mu(t)$ is close to $\mu\str(t)$.  

Note that $\eta(t)$ depends on $\lambda\str(t)$ and $\mu\str(t)$ through~\eqref{eq:defn_of_eta} so it becomes undefined if $\vaa\str(t)=0$.  Accordingly, let $\eta(t)$ be defined only on all time intervals where $\vaa\str(t)\neq0$ and assume that each of these intervals has nonzero measure.  Consider $n\str(t) = \tan(\tfrac{1}{4}\theta\str(\|\vaa\str(t)\|/r))$.  Whenever $\|\vaa\str(t)\|\leq \sca_{2\pi} r$, we have $\theta\str(\|\vaa\str(t)\|/r)=2\pi$ so $n\str(t)=\infty$.  Therefore, on any interval where $\vaa\str(t)=0$, the desired attitude set $n\str(t)\bar\BB^3$ is already expanded to include all of $\RR^3$.  This fact will be seen to enable the controller to be independent of $\lambda\str(t)$ and $\mu\str(t)$ on any interval where the target thrust can go to zero.

\subsubsection{Constructing a Thrust Feedback Law} \label{subsec:u1_law}
Having identified a set of MRPs that enables~\eqref{eq:geometry_1}, we now determine a compatible control law for $u_1(t)$.  The previous work~\cite{willebeek_lemair_practical_2026} addressed the $U_1=0$ case with the parsimonious rule $u_1(t)=\|\vaa\str(t)\|$.  In this work, we generalize the previous approach in two ways.  First, we introduce a control law that accommodates the $U_1>0$ case and simplifies to $u_1(t)=\|\vaa\str(t)\|$ when $U_1=0$.  Second, we introduce a further modification that accounts for the alignment of $\lambda(t)$ and $\lambda\str(t)$ by attenuating $u_1(t)$ when the two unit vectors are not aligned.  

\begin{figure}[t!]
    \centering
    \begin{minipage}{0.5\textwidth}
        \begin{subfigure}[t]{0.45\textwidth}
            \centering
            \includegraphics[trim={0.0cm 0cm 0.0cm -0.04cm},clip,width=0.95\textwidth]{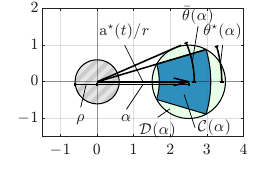}
            \caption{$\setC(\sca)$ for $\sca=2.5$}
            \label{fig:annulus_a}
        \end{subfigure}%
        ~
        \begin{subfigure}[t]{0.45\textwidth}
            \centering
            \includegraphics[trim={0.0cm 0cm 0.0cm -0.04cm},clip,width=0.95\textwidth]{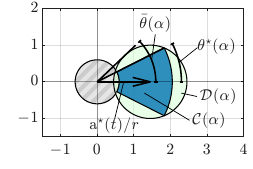}
            \caption{$\setC(\sca)$ for $\sca=1.45$}
            \label{fig:annulus_b}
        \end{subfigure}
    \end{minipage}
    \begin{minipage}{0.5\textwidth}
        \begin{subfigure}[t]{0.45\textwidth}
            \centering
            \includegraphics[width=0.95\textwidth]{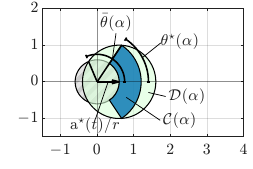}
            \caption{$\setC(\sca)$ for $\sca=0.6$}
            \label{fig:annulus_c}
        \end{subfigure}%
        ~
        \begin{subfigure}[t]{0.45\textwidth}
            \centering
            \includegraphics[width=0.95\textwidth]{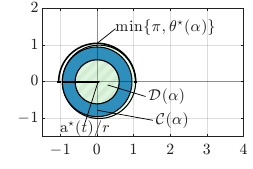}
            \caption{$\setC(\sca)$ for $\sca=0.05$}
            \label{fig:annulus_d}
        \end{subfigure}
    \end{minipage}
    \caption{Schematic of configurations of $\setC(\sca)$ for $\rho=0.6$.}
    \label{fig:annulus_schematic}
    \vspace{-0.4cm}
\end{figure}

The control laws for $u_1(t)$ can be understood through the lens of the following construction, illustrated in Fig.~\ref{fig:annulus_schematic}.  Suppose $\sca$ is large enough that $\setD(\sca)$ and ${\rm int}\rho\BB^2$ are disjoint, as shown in Fig.~\ref{fig:annulus_a}. In this case, the line $L(\theta\str(\sca))$ intersects $\partial\setD(\sca)$ at two points.  Sweeping this line segment $2\theta\str(\sca)$ radians clockwise around the origin yields the annulus section 
\begin{equation} \label{eq:defn_of_C0}
    \begin{aligned}
        \setC_0(\sca) &= \{\ell \, [\cos\phi \,\, \sin\phi]\tt \,|\, \phi\in[-\theta\str(\sca),\theta\str(\sca)], \\&\qquad \ell\in [l_-(\sca,\theta\str(\sca)),l_+(\sca,\theta\str(\sca))]\}
    \end{aligned}
\end{equation}
The annulus $\setC_0(\sca)$ lies entirely within $\setD(\sca)\setminus{\rm int}\rho\BB^2$ so it represents a set of $u_1(t),\lambda(t)$ pairs for which $u_1(t)\lambda(t)\in\calA(t)$ holds.  Consider slightly smaller values of $\sca$.  When $\setD(\sca)$ and ${\rm int}\rho\BB^2$ first intersect, as in Fig.~\ref{fig:annulus_b}, $\setC_0(\sca)$ still lies entirely within $\setD(\sca)\setminus{\rm int}\rho\BB^2$.  For slightly smaller values of $\sca$, as in Fig.~\ref{fig:annulus_c}, the value of $l_-(\sca,\theta\str(\sca))$ drops below $\rho$.  Accordingly, $\setC_0(\sca)$ no longer represents an annulus lying within $\setD(\sca)\setminus{\rm int}\rho\BB^2$.  This can be remedied by replacing $l_-(\sca,\theta\str(\sca))$ with $\max\{\rho,l_-(\sca,\theta\str(\sca))\}$ in~\eqref{eq:defn_of_C0}.  For even smaller values of $\sca$, the value of $\bar\theta(\sca)$ exceeds $\pi$.  In this case, $l_+(\sca,\theta\str(\sca)) > l_+(\sca,0) = 1-\sca$ so part of $\setC_0(\sca)$ lies outside of $\setD(\sca)$.  This can be addressed by replacing $l_+(\sca,\theta\str(\sca))$ by $l_+(\sca,\min\{\pi,\theta\str(\sca)\})=l\str_+(\sca)$ in~\eqref{eq:defn_of_C0}.  These two modifications yield
\begin{equation} \label{eq:defn_of_C} 
    \begin{aligned}
        \setC(\sca) &= \{\ell \, [\cos\phi \,\, \sin\phi]\tt \,|\, \phi\in[-\theta\str(\sca),\theta\str(\sca)], \\&\qquad \ell\in [\max\{\rho,l_-(\sca,\theta\str(\sca))\},l\str_+(\sca)]\}
    \end{aligned}
\end{equation}
which is the set illustrated in blue in Fig.~\ref{fig:annulus_schematic}.  For all $\sca\geq0$, $\setC(\sca)$ lies entirely within $\setD(\sca)\setminus{\rm int}\rho\BB^2$, and so it represents a set of $u_1(t),\lambda(t)$ pairs for which $u_1(t)\lambda(t)\in\calA(t)$ holds.  

For illustration purposes, $\setC(\sca)$ uses $\phi\in[-\theta\str(\sca),\theta\str(\sca)]$ rather than $\phi\in[0,\theta\str(\sca)]$, as is done in the rest of the paper.  This graphical choice was made to enable Fig.~\ref{fig:annulus_schematic} to suggest the geometry of the set corresponding to $\setC(\sca)$ in the main, three-dimensional model.  It does not impact the subsequent analysis, which returns to $\phi\in[0,\theta\str(\sca)]$.

Leveraging~\eqref{eq:defn_of_C} enables the relatively straightforward design condition formalized in the following.  Suppose that $\phi\in[0,\theta\str(\sca)]$.  Then a sufficient condition for $\ell \, [\cos\phi \,\, \sin\phi]\tt\in\setC(\sca)\subset \setD(\sca)\setminus{\rm int}\rho\BB^2$ is
\begin{equation}
    \ell \in [\max\{\rho,\sca\},\max\{\tfrac{1+\rho}{2},\sca\}] \label{eq:sufficient_for_C}
\end{equation}
where we have used~\eqref{eq:optional_l_star_condition_2} and the fact that $l_-(\sca,\phi)\leq \sca$ for all $(\sca,\phi)\in\dom l_- = \{(\alpha,\phi)\in\RR_{\geq0}\,|\,\alpha\sin\phi\leq 1\}$.

\begin{lemma} \label{lemma:kappa_circ}
    Define the radially symmetric (i.e., independent of $\phi$) schedule $\kappa_{\circ}:\RR_{\geq0}\to[\rho,\infty)$ by
    \begin{equation}
        \kappa_{\circ}(\sca)=\rho + Z(\sca;\alpha_0,\alpha_1),
        \label{eq:defn_of_kappa_circ}
    \end{equation}
    with $\alpha_0=\max\{0,\tfrac{3\rho-1}{2}\}$ and $\alpha_1=\min\{2\rho,\tfrac{1+\rho}{2}\}$. Then
    \begin{equation}
        \kappa_{\circ}(\sca) \in [\max\{\rho,l_-(\sca,\theta\str(\sca))\},l\str_+(\sca)],
        \label{eq:kappa_circ_is_good}
    \end{equation}
    for all $\sca\geq0$.
\end{lemma}

\begin{proof}
    From the definition of $Z$ in Appendix~\ref{app:zeta}: 1) $\kappa_{\circ}$ is nondecreasing, 2) $\kappa_{\circ}(\sca)=\sca$ for $\sca\geq\alpha_1$, and 3) $\kappa_{\circ}$ is a convex function.  From 1), it follows that $\kappa_{\circ}(\sca)\geq \kappa_{\circ}(0)=\rho$ for all $\sca\geq0$. From 2), the unique solution to $\kappa_{\circ}(\sca)=\tfrac{1+\rho}{2}\leq\alpha_1$ is $\sca=\tfrac{1+\rho}{2}$.  This implies that $\kappa_{\circ}(\sca)<\tfrac{1+\rho}{2}$ for $\sca<\tfrac{1+\rho}{2}$ and $\kappa_{\circ}(\sca)=\sca$ for $\sca\geq\tfrac{1+\rho}{2}$.  From 3), it follows that the epigraph of $\kappa_{\circ}$, $\epi \kappa_{\circ}$, is a convex set.  The line $\{(\sca,\sca)|\sca\in\RR\}$ is a supporting hyperplane of $\epi \kappa_{\circ}$, so $\kappa_{\circ}(\sca)\geq \sca$ for all $\sca\geq0$ according to the convexity of $\epi \kappa_{\circ}$.  Assembling all of the noted inequalities verifies~\eqref{eq:sufficient_for_C}, which implies~\eqref{eq:kappa_circ_is_good}.  
\end{proof}

An immediate corollary to Lemma~\ref{lemma:kappa_circ} is the following feedback law for $u_1(t)$.

\begin{corollary}
    If $\vaa\str(t)\neq0$ and $\theta(t)\in\Theta\str(t)$, then the radially symmetric feedback law
    \begin{equation}
        u_1(t) = r \, \kappa_\circ(\|\vaa\str(t)\|/r) \label{eq:kappa_circ_feedback_law}
    \end{equation}
    ensures $u_1(t)\lambda(t)\in\calA(t)$.  
\end{corollary}

\begin{remark}
    Consider~\eqref{eq:kappa_circ_feedback_law} in the case of $U_1>0$.  The feedback law ensures $u_1(t)=\|\vaa\str(t)\|$ whenever $\|\vaa\str(t)\|\geq\min\{2U_1,\tfrac{r+U_1}{2}\}$.  As $U_1$ shrinks toward zero, $\min\{2U_1,\tfrac{r+U_1}{2}\}$ likewise shrinks to zero.  For $U_1=0$, therefore,~\eqref{eq:kappa_circ_feedback_law} simplifies to $u_1(t)=\|\vaa\str(t)\|$.  The approach~\eqref{eq:kappa_circ_feedback_law} to accommodate $U_1>0$ is, therefore, a bona fide generalization of the approach developed in~\cite{willebeek_lemair_practical_2026}.
\end{remark}

The radial symmetry of $\kappa_\circ$ can be viewed as a deficiency: setting $u_1(t) = r\,\kappa_\circ(\|\vaa\str(t)\|/r)$ allows for the possibility of applying large thrusts in the wrong direction if $\lambda(t)$ and $\lambda\str(t)$ are significantly different.  The following corollary to Lemma~\ref{lemma:kappa_circ} provides an actuation law that is not radially symmetric.

\begin{corollary} \label{corollary:kappa}
    Define $\kappa:\RR_{\geq0}\times[0,\pi]\to[\rho,\infty)$ by 
    \begin{equation}
        \begin{aligned}
            \kappa(\sca,\phi) &= \kappa_{\circ}(\sca)(1-\zeta(\phi-\theta\str(\sca);\delta_0, \delta_1)) \\&\qquad + \rho \zeta(\phi-\theta\str(\sca);\delta_0, \delta_1)
        \end{aligned}
        \label{eq:defn_of_kappa}
    \end{equation}
    for some $\delta_1>\delta_0>0$. Then for all $\sca\in\RR_{\geq0}$, 
    \begin{equation}
        \kappa(\sca,\phi) = \kappa_{\circ}(\sca)\in [\max\{\rho,l_-(\sca,\theta\str(\sca))\},l\str_+(\sca)]
        \label{eq:kappa_is_good}
    \end{equation}
    whenever $\phi\in [0,\theta\str(\sca)]$.  
\end{corollary}

\begin{corollary}
    If $\vaa\str(t)\neq0$, then the feedback law
    \begin{equation}
        u_1(t) = r \, \kappa(\|\vaa\str(t)\|/r,\theta(t)) \label{eq:u1_prescription}
    \end{equation}
    ensures $u_1(t)\lambda(t)\in\calA(t)$ whenever $\theta(t)\in\Theta\str(t)$.  
\end{corollary}

One may observe that $\kappa(\sca,\phi)=\rho$ whenever $\sca$ approaches zero regardless of the value of $\phi$: when $\sca\leq\alpha_0$, $\kappa$ becomes a convex combination of $\kappa_{\circ}(\sca)=\rho$ with $\rho$.  We make use of this fact to extend the feedback law~\eqref{eq:u1_prescription} to include 
\begin{equation}
    \text{$\vaa\str(t)=0$ (i.e., $\theta(t)$ undefined)}  \quad \Rightarrow \quad u_1(t) = U_1.
\end{equation}
This extension ensures that $u_1(t)$ is well-defined during the free-fall singularity where $\theta(t)$ is undefined. For the remainder of the analysis, we take $u_1(t)$ to be given by~\eqref{eq:u1_prescription}.

\begin{remark}
    The choice of how to account for the alignment of $\lambda(t)$ and $\lambda\str(t)$ in the feedback law for $u_1(t)$ is an area of significant design freedom.  In simulation, different design choices have caused dramatically different qualities of transient performance.  While desirable transient behavior has been observed for~\eqref{eq:defn_of_kappa}, the transient performance degrades if $\delta_0$ and $\delta_1$ are selected to be too close to zero.  Any radially asymmetric thrust law may be used in place of~\eqref{eq:defn_of_kappa} in the following design so long as $\kappa(\sca,\phi)\in[\rho,\kappa_\circ(\sca)]$ for all $\phi$ and so long as $\kappa(\sca,\phi)=\kappa_\circ(\sca)$ for all $\phi\in[0,\theta\str(\sca)+\delta_0]$ for some finite $\delta_0>0$ that is independent of $\sca$.  
\end{remark}

\subsubsection{Designing a Lyapunov Function}

To guarantee convergence of $\vaa(t)=\lambda(t)u_1(t)$ to $\setA(t)$ through smooth feedback, we first construct a Lyapunov function for use in a backstepping control design process discussed in Section~\ref{subsubsec:backstepping}.  MRP-based attitude controllers can exhibit unwinding.  This unwinding makes global asymptotic stability (GAS) of $\setA(t)$ impossible --- stability in the sense of Lyapunov fails since it is possible for $\vaa(t)$ to begin close to the ball $\setA(t)$, move away as $\eta(t)$ unwinds, and then return as $t\to\infty$.  Instead of proving GAS, the set $\setA(t)$ is shown to be globally attractive.  The proof relies on the development of a candidate Lyapunov function $V_{\vaa}(t,\eta)$ for which 
\begin{equation}
    \|\vaa(t)\|_{\setA(t)}\leq \chi\bigl(V_{\vaa}(t,\eta(t))\bigr)
    \label{eq:overestimate_goal}
\end{equation}
for some $\chi\in\calK_{\infty}$.  

Whenever $\vaa(t)\not\in\setA(t)$, the distance from $\vaa(t)$ to $\setA(t)$ is
\begin{equation}
    \begin{aligned}
        \|\vaa(t)\|_{\setA(t)} &= \Bigl\|r\kappa(\|\vaa\str(t)\|/r,\theta(t))\lambda(t)-\vaa\str(t)\Bigr\|-r.
    \end{aligned}
\end{equation}
The following lemma, whose proof is deferred to Appendix~\ref{appsec:angle_bound_on_dist}, provides an intermediate upper bound.  

\begin{lemma} \label{lemma:angle_bound_on_dist}
    The distance from $\vaa(t)$ to $\setA(t)$ satisfies 
    \begin{equation}
        \|\vaa(t)\|_{\setA(t)}\leq \ell_{\vaa} \|\vaa\str(t)\|\max\{0,\theta(t)-\theta\str(\|\vaa\str(t)\|/r)\}
        \label{eq:angle_bound_on_dist}
    \end{equation}
    for some $\ell_{\vaa}>0$ whenever $\vaa\str(t)\neq0$ and is zero otherwise.
\end{lemma}

The arclength of the MRP rotation $\eta$ is $4\arctan(\|\eta\|)$, which takes values in $[0,2\pi]$ since MRPs are a double covering of $SO(3)$.  The following corollary uses this expression and the fact that $a-b\geq\arctan(a)-\arctan(b)$ whenever $a>b$ to relax the bound of Lemma~\ref{lemma:angle_bound_on_dist} in terms of MRPs.  

\begin{corollary} \label{corollary:MRP_bound_on_dist}
    The distance from $\vaa(t)$ to $\setA(t)$ is bounded by
    \begin{equation}
        \|\vaa(t)\|_{\setA(t)}\leq 4\ell_{\vaa} \|\vaa\str(t)\|\max\{0,\|\eta(t)\|-n\str(t)\}
        \label{eq:MRP_bound_on_dist}
    \end{equation}
    for some $\ell_{\vaa}>0$ whenever $\vaa\str(t)\neq0$ and is zero otherwise.
\end{corollary}

From Corollary~\ref{corollary:MRP_bound_on_dist} we can develop a candidate Lyapunov function $V_{\vaa}(t,\eta)$ that satisfies~\eqref{eq:overestimate_goal}.  Noting that the inverse of a class $\calK_{\infty}$ function is also a class $\calK_{\infty}$ function, define the function $\chi\in\calK_{\infty}$ by its inverse
\begin{equation}
    \chi^{-1}(s) = Z(\tfrac{1}{4\ell_{\vaa}}s;0,\Delta_{\vaa})
    \label{eq:defn_of_chi}
\end{equation}
for some $\Delta_{\vaa}>0$. Next, define $V_{\vaa}:\RR_{\geq0}\times\bar\RR^3\to\RR_{\geq0}$ by
\begin{equation}
    V_{\vaa}(t,\eta) = Z(\|\vaa\str(t)\| (\|\eta\|-n\str(t));0,\Delta_{\vaa}). \label{eq:defn_of_V_eta}
\end{equation}
The following corollary is then the direct result of~\eqref{eq:defn_of_chi},~\eqref{eq:defn_of_V_eta}, and Corollary~\ref{corollary:MRP_bound_on_dist}.

\begin{corollary} \label{corollary:lyapunov_bound_on_dist}
    The distance from $\vaa(t)$ to $\setA(t)$ satisfies 
    \begin{equation}
        \|\vaa(t)\|_{\setA(t)}\leq \chi(V_{\vaa}(t,\eta(t)))
        \label{eq:lyapunov_bound_on_dist}
    \end{equation}
    for some $\chi\in\calK_{\infty}$ whenever $\vaa\str(t)\neq0$ and is zero otherwise.
\end{corollary}

\subsubsection{Backstepping Analysis} \label{subsubsec:backstepping}
Consider the candidate Lyapunov function $V_{\vaa}(t,\eta)$.  Omitting arguments of time and simplifying, the derivative of $V_{\vaa}$ along trajectories is 
\begin{equation}
    \begin{aligned}
        \dot V_{\vaa}
        &= \zeta(\|\vaa\str\|(\|\eta\|-n\str);0,\Delta_{\vaa}) \Bigl(\lambda\str{}\tt\dot\vaa\str (\|\eta\|-n\str)
        \\&
        + \|\vaa\str\|\Bigl(\frac{1+\|\eta\|^2}{4}\frac{\eta\tt}{\|\eta\|}\bigl(\omega(t)-R(t)\tt\!\bigl(\hat\lambda\str(t)\tfrac{\dot\vaa\str(t)}{\|\vaa\str(t)\|} 
        \\&\qquad\qquad\qquad
        + \lambda\str(t) \Omega_\mu(t)\bigr)\bigr) -\dot n\str \Bigr) \Bigr),
    \end{aligned} \label{eq:ddt_V_eta_1}
\end{equation}
where we have used the fact that $\eta\tt T(\eta) = \tfrac{1}{4}(1+\|\eta\|^2)\eta$.

Define the target angular velocity $\omega\str:\RR_{\geq0}\times\bar\RR^3\to\RR^3$ by
\begin{align}
    &\omega\str(t,\eta) 
    = 
    \biggl(- \frac{4(\eta/\|\eta\|)}{1+\|\eta\|^2} \frac{\Omega_{\vaa}(t,\eta)}{\|\vaa\str(t)\|} 
    + E^{-1}(\eta)\tt\e_3\Omega_\mu(t)
    \notag\\&
    +\!\Bigl(\!E^{-\!1}\!(\eta)\!{}\tt \! R\str\!(t)\!{}\tt\hat\lambda\str\!(t)-\frac{4(\eta/\|\eta\|)}{1+\|\eta\|^2}(\!\|\eta\|\!-\!n\str\!(t)\!)\lambda\str\!(t)\!{}\tt\!\Bigr)\frac{\dot\vaa\str(t)}{\|\vaa\str(t)\|}
    \notag\\&
    + \frac{4(\eta/\|\eta\|)}{1+\|\eta\|^2}  \Bigl(\dot n\str(t)\bigl(1-\zeta(\dot n\str(t);0,\Delta_{\dot n\str})\bigr)\Bigr)\biggr)\zeta(\frac{\|\eta\|}{n\str(t)};\varrho,1)
    \notag\\&
    - \frac{4 \gain_{\eta}\eta}{1+\|\eta\|^2} \zeta(\|\vaa\str(t)\|;\vaa_0,\vaa_1)
    \label{eq:defn_of_omega_star}
\end{align} 
whenever $\vaa\str(t)\neq0$ with $\omega\str(t,\eta)=0$ otherwise, for some $\Delta_{\dot n\str}\in(0,\infty)$, $\varrho\in(0,1)$, $\gain_{\eta}\in(0,\infty)$, $\vaa_0\in(0,\infty)$, and $\vaa_1\in(\vaa_0,\infty)$, where $\Omega_{\vaa}:\RR_{\geq0}\times\bar\RR^3\to\RR$ is yet to be specified. Note that we use $\frac{\eta}{\|\eta\|}\zeta(\frac{\|\eta\|}{n\str(t)};\varrho,1) = 0$ when $\|\eta\|\leq \varrho n\str(t)$. The first term in~\eqref{eq:defn_of_omega_star}, involving $\Omega_{\vaa}$, is the one primarily responsible for driving $V_{\vaa}$ to zero in the subsequent Lyapunov analysis. The second and third terms on the right of~\eqref{eq:defn_of_omega_star}, involving $\Omega_\mu(t)$ and $\dot\vaa\str(t)$, are feedforward terms responsible for driving the gimbal attitude vectors $\lambda(t),\mu(t)$ toward the target values $\lambda\str(t),\mu\str(t)$ and for speeding up the convergence of the rotational coordinate $\eta(t)$ to $N\str(t)$ whenever the magnitude of $\vaa\str(t)$ is growing rapidly.  The fourth term in~\eqref{eq:defn_of_omega_star}, involving $\dot n\str(t)$, is responsible for keeping $\eta(t)$ inside of $N\str(t)$.   The fifth and final term in~\eqref{eq:defn_of_omega_star} keeps $\eta(t)$ small once it reaches $N\str(t)$.  

In conventional backstepping designs for multirotors~\cite{lee_geometric_2010,lee_control_2011,lee_geometric_2011,lee_nonlinear_2013,gamagedara_geometric_2019,martins_global_2021,martins_global_2024,flores_robust_2025}, the angular velocity feedforward terms present a singularity at $\vaa\str=0$.  The proposed design does not suffer this issue.  Suppose that $\vaa\str\to0$.  Then before $\vaa\str=0$, the target acceleration must pass into the neighborhood of $0$ where $\|\vaa\str\|\in(0, \sca_{2\pi}r]$.  Within this interval, $N\str(t)=\RR^3$ and $n\str(t)=\infty$.  It follows that $\|\eta(t)\|/n\str(t)\leq\varrho$ so $\zeta(\|\eta(t)\|/n\str(t);\varrho,1)=0$.  Hence, the first four terms in~\eqref{eq:defn_of_omega_star} smoothly transition to zero whenever $\|\vaa\str\|$ enters $[0,\sca_{2\pi}r]$.  Colloquially, these four control actions are \textit{turned off} whenever $\|\vaa\str\|\leq \sca_{2\pi}r$ which ensures that the free-fall singularity does not cause $\omega\str(t,\eta(t))$ to become discontinuous along any trajectory.  

The Lyapunov function is specifically designed to accommodate the four terms being smoothly switched on and off.  The switching has no direct impact on $\dot{V}_{\vaa}$ as
\begin{equation}
    \begin{aligned}
        \zeta(\|\eta\|/n\str&;\varrho,1) \, \zeta(\|\vaa\str\|(\|\eta\| \!-\! n\str);0,\Delta_{\vaa}) \\&= \zeta(\|\vaa\str\|(\|\eta\| \!-\! n\str);0,\Delta_{\vaa}) 
    \end{aligned} \label{eq:vanishing_step}
\end{equation}
since $\zeta(\|\eta\|/n\str;\varrho,1)=1$ whenever $\zeta(\|\vaa\str\|(\|\eta\| \!-\! n\str);0,\Delta_{\vaa})$ is nonzero.  The identity~\eqref{eq:vanishing_step} ensures the first four terms in~\eqref{eq:defn_of_omega_star} drive $\dot{V}_{\vaa}$ as if the switch $\zeta(\|\eta\|/n\str;\varrho,1)$ were absent.

Adding and subtracting $\omega\str$ on the right hand side of~\eqref{eq:ddt_V_eta_1} then simplifying reveals
\begin{align}
    &\dot V_{\vaa} = - \Bigl(\Omega_{\vaa} + \dot  n\str\,\|\vaa\str\|\,\zeta(\dot  n\str;0,\Delta_{\dot n\str}) 
    \label{eq:ddt_V_eta_2}
    \\ & \,\,\,\,\,
    + \gain_\eta \|\eta\| \|\vaa\str\| \, \zeta(\|\vaa\str\|;\vaa_0,\vaa_1)  \Bigr)  \zeta(\|\vaa\str\|(\|\eta\|\!-\!n\str);0,\Delta_{\vaa}) 
    \notag
    \\& \,\,\,\,\,
    + \bigl(\tfrac{1+\|\eta\|^2}{4}\bigr) \|\vaa\str\| \zeta(\|\vaa\str\|(\|\eta\|\!-\!n\str);0,\Delta_{\vaa}) 
    \tfrac{\eta\tt}{\|\eta\|}\bigl( \omega - \omega\str \bigr). \notag
\end{align} 
To ensure $V_{\vaa}\to0$, take $\gain_{\vaa} >0$ and prescribe 
\begin{equation}
    \hspace{-0.2cm}
    \Omega_{\vaa}(t,\eta) = \left\{ 
    \matl{
    \gain_{\vaa}  \frac{Z(\|\vaa\str(t)\|(\|\eta\| - n\str(t));0,\Delta_{\vaa})}{\zeta(\|\vaa\str(t)\|(\|\eta\| - n\str(t));0,\Delta_{\vaa})}, & \!\!\!\!V_{\vaa}(t,\eta)>0 \\
    0, & \!\!\!\!\text{otherwise}
    }
    \right., \hspace{-0.1cm} \label{eq:defn_of_Omega_eta}
\end{equation}
which is smooth and nonnegative.  With this choice, the term involving $\Omega_{\vaa}$ in~\eqref{eq:ddt_V_eta_2} becomes
\begin{equation}
    -\Omega_{\vaa} \, \zeta(\|\vaa\str\|(\|\eta\|\!-\! n\str);0,\Delta_{\vaa})  = -\gain_{\vaa}  V_{\vaa}. 
\end{equation}

Expanding the scope of the backstepping design to the entire inner-loop, consider the candidate Lyapunov function
\begin{equation}
    V(t,\eta,\omega) = V_{\vaa}(t,\eta) + \underbrace{\frac{1}{2} p_{\omega} \|\omega-\omega\str(t,\eta)\|^2}_{\eqqcolon \, V_\omega(t,\eta,\omega)},
\end{equation}
where $p_\omega>0$.  Omitting arguments of time, the derivative of $V$ along trajectories is 
\begin{align}
    \dot V 
    &= \dot V_{\vaa} + p_{\omega} (\omega-\omega\str)\tt(u_2 - \dot\omega\str)
    \label{eq:ddt_V_1} \\ 
    &= -\gain_{\vaa} V_{\vaa} - \Bigl(\dot n\str\|\vaa\str\|\zeta(\dot  n\str;0,\Delta_{\dot n\str}) 
    \label{eq:ddt_V_3} 
    \\ & \quad \,\,
    +\! \gain_\eta \|\eta\| \|\vaa\str\|  \zeta(\|\vaa\str\|;\vaa_0,\vaa_1)\!  \Bigr)  \zeta(\|\vaa\str\|(\|\eta\|\!-\!n\str);\!0,\!\Delta_{\vaa}) 
    \notag \\ & \quad 
    + p_{\omega} (\omega-\omega\str)\tt\Bigl(u_2 - \dot\omega\str
    \notag
    \\ & \quad  \,\,
    + p_{\omega}^{-1} \tfrac{\eta}{\|\eta\|}\bigl(\tfrac{1+\|\eta\|^2}{4}\bigr) \|\vaa\str\| \zeta(\|\vaa\str\|(\|\eta\|\!-\!n\str);0,\Delta_{\vaa}) \Bigr).
    \notag
\end{align}
Set $\gain_\omega >0$ and prescribe
\begin{equation}
    \begin{aligned}
        u_2(t) &= \dot\omega\str(t,\eta(t),\omega(t)) - \tfrac{1}{2} \gain_{\omega} (\omega-\omega\str) 
        \\ & \quad 
        -p_{\omega}^{-1}\tfrac{\eta(t)}{\|\eta(t)\|}\bigl(\tfrac{1+\|\eta(t)\|^2}{4}\bigr)\|\vaa\str(t)\| 
        \\ & \qquad 
        \cdot \, \zeta(\|\vaa\str(t)\|(\|\eta(t)\|\!-\! n\str(t));0,\Delta_{\vaa}).
    \end{aligned} \label{eq:u2_prescription}
\end{equation}
where we take the third term on the right hand side to be zero whenever either $\eta(t)=0$ or $\vaa\str(t)=0$.  Also note that $\dot\omega\str=0$ whenever $\vaa\str(t)=0$.  Combining~\eqref{eq:ddt_V_3} with~\eqref{eq:u2_prescription} and simplifying reveals
\begin{equation}
    \dot V \leq -\gain_{\vaa} V_{\vaa} - \gain_{\omega}V_\omega \leq -\gain V \label{eq:sufficient_for_attractivity}
\end{equation}
where $\gain=\min\{\gain_{\vaa} ,\gain_{\omega}\}$.  Applying the Comparison Lemma to this result yields the following theorem.  

\begin{theorem} \label{thrm:a_to_A}
    Under the action of the control law defined in~\eqref{eq:u1_prescription} and~\eqref{eq:u2_prescription}, $\setA(t)$ globally attracts $\vaa(t)=\lambda(t)u_1(t)$.  
\end{theorem}

\begin{proof}
    Equation~\eqref{eq:sufficient_for_attractivity} and the Comparison Lemma~\cite{khalil_nonlinear_2002} together imply that $V$ decays exponentially to zero:
    \begin{equation}
        V(t,\eta(t),\omega(t))\leq V(T,\eta(T),\omega(T)) e^{-k(t-T)}
        \label{eq:decaying_bound_on_V}
    \end{equation}
    for all $T\geq0$.  Recalling~\eqref{eq:overestimate_goal} and that $V_{\vaa}\leq V$, 
    \begin{equation}
        \|\vaa(t)\|_{\setA(t)}\leq \chi\bigl(V(T,\eta(T),\omega(T)) e^{-k(t-T)}\bigr) 
        \label{eq:decaying_bound_on_set_dist}
    \end{equation}
    for all $T\geq0$.  Selecting $T=0$ and taking $t\to\infty$ shows that $\|\vaa(t)\|_{\setA(t)}\to0$, so $\vaa(t)\to\setA(t)$ globally.
\end{proof}

\subsubsection{MRP Switching Condition} \label{subsec:switching_condition}

Before analyzing the behavior of the outer-loop system in light of the established results, this section introduces the MRP switching condition for the control law.  Formal notation for hybrid systems~\cite{goebel_hybrid_2009} is omitted for brevity.  In the following, $\eta$ denotes the value of the attitude error before the switch and $\eta^+$ denotes the value after the switch.  The switching condition is defined as
\begin{subequations} \label{eq:switching_condition}
    \begin{align}
        V(t,\mathrm{sh}(\eta),\omega) + \varepsilon\leq V(t,\eta,\omega) \, &\Rightarrow \, \eta^+ \!= \mathrm{sh}(\eta)
        \label{eq:lyapunov_jump}
        \\
        \hspace{-0.2cm}
        \|\vaa\str\|\!\in\!(0,\min\{\sca_{2\pi}r,\vaa_0\}) \, \text{and} \, \|\eta\|\!>\!1 \, &\Rightarrow \, \eta^+ \!= E(\tilde R)
        \label{eq:free_fall_jump}
    \end{align}
\end{subequations}
for some $\varepsilon>0$, where we have omitted arguments of time for clarity.  Whenever the switching condition is inactive, $\eta(t)$ evolves according to~\eqref{eq:defn_of_ddt_eta}.  

The switching condition~\eqref{eq:lyapunov_jump} is designed to prevent unwinding.  Suppose that $\|\eta\|>\|\mathrm{sh}(\eta)\|$ and that $\omega$ is small.  This is the classic setup for unwinding.  In this case, $V(t,\eta,\omega)$ will be larger than $V(t,\mathrm{sh}(\eta),\omega)$, so~\eqref{eq:lyapunov_jump} switches from $\eta$ to $\mathrm{sh}(\eta)$ and prevents unwinding.  Next, suppose that $\|\eta\|>\|\mathrm{sh}(\eta)\|$ and that $\omega$ is \textit{large}.  Further suppose that $\omega$ is close to $\omega\str(t,\eta)$ and far from $\omega\str(t,\mathrm{sh}(\eta))$.  In this case, the condition for~\eqref{eq:lyapunov_jump} may not be fulfilled.  If it is not fulfilled, then the system may unwind.  This unwinding is actually desirable since the angular velocity $\omega$ will have carried the system through the target motion with minimal control effort.  

The switching condition~\eqref{eq:free_fall_jump} is designed to ensure desirable performance near the free-fall singularity.  Whenever $\|\vaa\str(t)\|$ is small enough that $N\str(t)=\RR^3$, switching $\eta(t)$ has no effect on the value of $V$ and no effect on the control action.  In this regime it is therefore reasonable to switch $\eta(t)$ according to~\eqref{eq:free_fall_jump} so that $\eta(t)$ is in the covering of $SO(3)$ close to the origin.  This switch ensures that $\eta(t)$ is well-defined immediately after any time where $\vaa\str(t)=0$, and it discourages unwinding after such events since the $E(\tilde R)$ covering of $SO(3)$ is the one close to the origin.  

Consider the consequences of~\eqref{eq:switching_condition} for the value of $V$ along trajectories.  Whenever $\eta$ is switched due to~\eqref{eq:lyapunov_jump}, the value of $V$ decreases by a finite increment $V^+\leq V-\varepsilon$.  Whenever $\eta$ is switched due to~\eqref{eq:free_fall_jump}, the value of $V$ is constant along trajectories $V^+=V$.  Therefore, applying~\eqref{eq:switching_condition} does not change the result of Theorem~\ref{thrm:a_to_A}  ---  $\vaa(t)$ is still asymptotically attracted to $\setA(t)$.

\subsection{Long-term Behavior of the Composite System} \label{subsec:long_term}

Consider the behavior of the inner- and outer-loop systems when they are connected to one another.  Set $\vaa\str(t) = \kappa_{ x\str}(t,e(t))$ and define $\Delta\vaa:\RR_{\geq0}\to\RR^3$ according to
\begin{equation}
    \Delta\vaa(t)= \vaa(t)-\vaa\str(t).
\end{equation}
It follows that the first row of~\eqref{eq:olsys} can be rewritten in terms of $e(t) =  x(t)- x\str(t)$ as
\begin{equation}
    \begin{aligned}
        \dot e(t) &= f_{ x\str}(t,e(t)) + \g_{ x\str}(t,e(t))\kappa_{ x\str}(t,e(t)) 
        \\
        &\qquad 
        + \g_{ x\str}(t,e(t)) \Delta\vaa(t).
    \end{aligned} \label{eq:error_outer_clsys}
\end{equation}
The ISS property~\eqref{eq:outer_clsys_is_pre_ISS} of the outer-loop system~\eqref{eq:surrogate_olsys} then implies that for any $T\in[0,\infty)$,  
\begin{align}
    \hspace{-0.2cm}
    \|e(t)\| \leq \beta(\|e(T)\|,t\!-\!T) + \gamma({\sup}_{\tau\geq T} \, \bar\g \|\Delta\vaa(\tau)\|) \hspace{-0.1cm} \label{eq:using_ISS}
\end{align}
for all $t\in [T,\infty)$, where $\bar\g=\sup_{t\in\RR_{\geq0},x\in\RR^n}\|\g(t, x)\|$. Theorem~\ref{thrm:a_to_A} provides a bound on $\|\Delta\vaa(t)\|\leq r+\|\vaa(t)\|_{\setA(t)}$.  This bound is applied to strengthen the ISS-type estimate~\eqref{eq:using_ISS} into global asymptotic stability (GAS) and ultimate boundedness.  

\begin{theorem} \label{thrm:asymptotic_stabilization}
    Under~\eqref{eq:u1_prescription},~\eqref{eq:u2_prescription}, for any $T\in[0,\infty)$ the ball 
    \begin{equation} \label{eq:stable_ball}
        \calB_T=\gamma\bigl(\bar\g r + \bar\g \chi(V(T,\eta(T),\omega(T)))\bigr)\bar\BB^n
    \end{equation}
    is GAS on the time interval $[T,\infty)$.
\end{theorem}

\begin{proof}
    Consider~\eqref{eq:using_ISS}.  From $\|\Delta\vaa(\tau)\|\leq r+\|\vaa(t)\|_{\setA(t)}$ and~\eqref{eq:decaying_bound_on_set_dist}, the second term on the right is bounded by
    \begin{align}
        \hspace{-0.2cm}
        \gamma({\sup}_{\tau\geq T} \, \bar\g \|\Delta\vaa(\tau)\|) 
        &\leq 
        \gamma(\bar\g r \!+\! {\sup}_{\tau\geq T} \, \bar\g \chi(V_T e^{-k(\tau\!-\!T)}))
        \hspace{-0.2cm}
        \notag
        \\&=
        \gamma(\bar\g r + \bar\g \chi(V_T))
    \end{align}
    for any $T\in[0,\infty)$, where $V_T=V(T,\eta(T),\omega(T))$.  
    Hence,
    \begin{align}
        \|e(t)\| \leq \beta(\|e(T)\|,t\!-\!T) +\gamma(\bar\g r + \bar\g \chi(V_T)) 
        \label{eq:did_not_expect_to_reference_this_TWICE_haha}
    \end{align}
    for all $t\in[T,\infty)$, for any $T\in[0,\infty)$.  Note that  
    \begin{equation}
        \|e(t)\|_{\calB_T}=\max\{0,\|e(t)\|-\gamma(\bar\g r + \bar\g \chi(V_T))\}.
    \end{equation}
    Then subtracting $\gamma(\bar\g r + \bar\g \chi(V_T))$ from both sides of~\eqref{eq:did_not_expect_to_reference_this_TWICE_haha} and using the fact that $0\leq\beta(\|e(T)\|,t\!-\!T)$, it follows that
    \begin{align}
        \|e(t)\|_{\calB_T} \leq \beta(\|e(T)\|,t\!-\!T) \label{eq:yet_another_reference}
    \end{align}
    for all $t\in[T,\infty)$, for any $T\in[0,\infty)$. The estimate~\eqref{eq:yet_another_reference} proves GAS of $\calB_T$ on the time interval $[T,\infty)$~\cite{chaillet_uniform_2008}.  
\end{proof}

\begin{theorem} \label{thrm:asymptotic_tracking}
    Under~\eqref{eq:u1_prescription},~\eqref{eq:u2_prescription}, the outer-loop error is ultimately bounded according to $\limsup{}_{t\to\infty} \|e(t)\| \leq \gamma(\bar\g r)$. 
\end{theorem}

\begin{proof}
    Consider~\eqref{eq:did_not_expect_to_reference_this_TWICE_haha}.  Taking the limit supremum as $t\to\infty$ drives the $\calK\calL$ term to zero, leaving
    \begin{equation}
        \hspace{-0.2cm}
        \limsup_{t\to\infty} \|e(t)\| 
        \leq \gamma(\bar\g r + \bar\g \chi(V_T))  \quad \forall \, T\in[0,\infty).
    \end{equation}
    Let $V_0=V(0,\eta(0),\omega(0))$.  Using $V_T\leq V_0 e^{-kT}$ from~\eqref{eq:decaying_bound_on_V},
    \begin{align}
        \limsup_{t\to\infty} \|e(t)\| 
        \leq \gamma\bigl(\bar\g r+\bar\g\chi(V_0 e^{-kT}) \bigr) \quad \forall \, T\in[0,\infty)
        \label{eq:ultimate_bound_with_T}
    \end{align}
    Since the right hand side of~\eqref{eq:ultimate_bound_with_T} is continuous in $T$ and it holds for all $T\in[0,\infty)$,~\eqref{eq:ultimate_bound_with_T} must hold as $T\to\infty$.  Hence, 
    \begin{align}
        \limsup_{t\to\infty} \|e(t)\| 
        \leq \lim_{T\to\infty}\gamma\bigl(\bar\g r+\bar\g \chi(V_0 e^{-kT}) \bigr) = \gamma(\bar\g r)
    \end{align}
    which is the claimed result.
\end{proof}

Theorems~\ref{thrm:asymptotic_stabilization} and~\ref{thrm:asymptotic_tracking} describe the transient and long-term behavior of the outer-loop system.  Theorem~\ref{thrm:asymptotic_stabilization} guarantees the error is asymptotically stabilized to a neighborhood of the origin, and Theorem~\ref{thrm:asymptotic_tracking} guarantees that the stabilized set eventually converges to within an ultimate bound.  Theorem~\ref{thrm:asymptotic_tracking} achieves the control objective~\eqref{eq:control_objective} with 
\begin{equation}
    \Delta = \gamma(\bar\g r).
\end{equation}
The size of $\Delta$ can be tuned to ensure practical utility by selecting an appropriate value of $r$, since $\gamma\in\calK_{\infty}$.

Theorem~\ref{thrm:no_ECT} established the nonexistence of ECT controllers for systems of the form~\eqref{eq:olsys}.  Theorem~\ref{thrm:asymptotic_tracking} provides a definitive positive answer to the question of existence of PCT controllers for the same class, formalized in the following corollary.  

\begin{corollary}
    Given Assumption~\ref{assmp:ISS_control_law}, there exists a PCT control law for all systems of the form~\eqref{eq:olsys} for any $\Delta>\gamma( \bar\g U_1)$. 
\end{corollary}

\begin{proof}
    Fix $\Delta>\gamma( \bar\g U_1)$ and select $r = \bar\g^{-1}\gamma^{-1}(\Delta)$.  This radius $r$ is compatible with the control law~\eqref{eq:u1_prescription},~\eqref{eq:u2_prescription} since it is greater than $U_1$.  Theorem~\ref{thrm:asymptotic_tracking} guarantees that applying~\eqref{eq:u1_prescription},~\eqref{eq:u2_prescription} with the selected $r$ ensures~\eqref{eq:control_objective}, the definition of PCT.  
\end{proof}

\section{Application to Multirotor UAVs} \label{sec:UAVs}

In this section, the theoretical results of Section~\ref{sec:controller_design} are applied to synthesize a practical complete position tracking controller for a multirotor UAV.  The position tracking controller is most appropriate for autonomous missions, and variants can be synthesized for other use cases.

The multirotor motion is described by the standard model
\begin{subequations} \label{eq:uav}
    \begin{align}
        \dot \x(t) &= \v(t)
        \label{eq:uav_position}
        \\
        \dot \v(t) &= g\e_3 + \frac{1}{m}\lambda(t) T(t)
        \label{eq:uav_velocity}
        \\
        \dot\lambda(t) &= (R_{IB}(t)\omega_B(t)) \times \lambda(t) 
        \label{eq:uav_pose_1}
        \\
        \dot\mu(t) &= (R_{IB}(t)\omega_B(t)) \times \mu(t) 
        \label{eq:uav_pose_2}
        \\
        \dot\omega_B(t) &= J^{-1}\pp{(J \omega_B(t))\times\omega_B(t) + \tau(t)}
        \label{eq:uav_angular_rate}
\end{align}
\end{subequations}
where $\x\in C^1(\RR_{\geq0},\RR^3)$ denotes position, $\v\in C^0(\RR_{\geq0},\RR^3)$ denotes velocity, $\lambda\in C^1(\RR_{\geq0},\SS^2)$ is the unit vector aligned to the vehicle's thrust, $\mu\in C^1(\RR_{\geq0},\SS^2)$ is the unit vector pointing out of the front of the vehicle, $R_{IB}\in C^1(\RR_{\geq0},SO(3))$ is the rotation matrix representation of the attitude $R_{IB}(t) = [\mu(t) \,\, -\lambda(t)\!\times\!\mu(t) \,\, -\lambda(t)]$, $\omega_B\in C^0(\RR_{\geq0},\RR^3)$ is the angular velocity of the vehicle, $T\in {\rm PWC}(\RR_{\geq0},[U_1,\infty))$ is the collective thrust input which is lower-bounded by $U_1\geq0$, and $\tau\in {\rm PWC}(\RR_{\geq0},\RR^3)$ is the torque input.  Additionally $g$ is the gravitational acceleration, $m$ is the vehicle's mass, and $J$ is the vehicle's inertia matrix.  

Toward putting~\eqref{eq:uav} in the form of~\eqref{eq:olsys}, apply the feedback 
\begin{equation}
    T(t) = u_1(t), \,\,\, \tau(t) = \omega_B(t)\!\times\!(J \omega_B(t)) + J R_{CB}\tt u_2(t)
\end{equation}
wherein $R_{CB} = \e_1\e_1\tt - \e_2\e_2\tt - \e_3\e_3\tt$ and introduce $R(t) = R_{IB}(t) R_{CB}\tt$ and $\omega(t) = R_{CB}\omega_B(t)$.  This results in
\begin{subequations} \label{eq:uav_2}
    \begin{align}
        \bbm{\dot\x(t) \\ \dot\v(t)} &= \bbm{\v(t)\\ ge_3} \!+\! \bbm{0\\ m^{-1}I} \lambda(t) u_1(t)
        \label{eq:uav_position_and_velocity}
        \\
        \dot\lambda(t) &= (R(t)\omega(t)) \times \lambda(t) 
        \label{eq:uav_gimbal_pose_1}
        \\
        \dot\mu(t) &= (R(t)\omega(t)) \times \mu(t) 
        \label{eq:uav_gimbal_pose_2}
        \\
        \dot\omega(t) &= u_2(t)
        \label{eq:uav_gimbal_rate}
\end{align}
\end{subequations}
which is of the form~\eqref{eq:olsys} with $x(t)=[\x(t)\tt \,\,\,\v(t)\tt]\tt$, $f(t,x(t))=[\v(t)\tt \,\,\, g\e_3\tt]\tt$, and $\g(t,x(t))=[0\,\,\,m^{-1}I]\tt$.

Equation~\eqref{eq:uav_position_and_velocity} is the description of the outer-loop system that will be used to synthesize the outer-loop controller $\kappa_{x\str}$.  In particular, replacing $\lambda(t) u_1(t)$ with $u_\vaa(t)$ and adding a disturbance term $\w(t)$, results in 
\begin{equation}
    \hspace{-1cm}\bbm{\dot\x(t) \\ \dot\v(t)} = \bbm{\v(t)\\ ge_3} \!+\! \bbm{0\\ m^{-1}I} u_\vaa(t) + \w(t)
    \label{eq:outer_position_olsys}
\end{equation}
which is in the form of~\eqref{eq:surrogate_olsys}.

For any smooth desired inertial position history $\x\str:\RR_{\geq0}\to\RR^3$, take $x\str(t) = [\x\str(t)\tt \,\,\,\dot\x\str(t)\tt]\tt$.  It is well known that a linear system $\dot x=Ax+B\w$ is ISS with respect to the disturbance $\w$ if and only if $A$ is Hurwitz~\cite{khalil_nonlinear_2002}.  This fact ensures that for any smooth desired outer-loop history $x\str$, the smooth control law $u_\vaa(t) = \kappa_{x\str}(t,x(t)-x\str(t))$ with
\begin{equation} \label{eq:outer_claw}
    \kappa_{x\str}(t,e) = m\ddot\x\str(t)-mg\e_3-[\gain_P I \,\,\, \gain_D I]e,
\end{equation}
wherein $\gain_P,\gain_D>0$, input-to-state stabilizes~\eqref{eq:outer_position_olsys} about $x\equiv x\str$ with respect to $\w$.  Along with~\eqref{eq:outer_claw}, we prescribe
\begin{equation} \label{eq:outer_claw_2}
    \nu(s) = \pp{\frac{\omega_{\nu}^2}{s^2+2\zeta_\nu\omega_{\nu}s+\omega_{\nu}^2}}(I-\e_3\e_3\tt)\v(s)
\end{equation}
where $s$ is the Laplace variable, $\nu(s)$ and $\v(s)$ are the Laplace transforms of $\nu(t)$ and $\v(t)$, and both $\omega_{\nu}$ and $\zeta_\nu$ are positive parameters.  The second-order low-pass filter in~\eqref{eq:outer_claw_2} ensures derivatives of $\nu(t)$ are available.  The design~\eqref{eq:outer_claw_2} ensures the vehicle's longitudinal axis, $\mu(t)=R_{IB}(t)\e_1$, is driven to align with the direction of motion over the ground, $(I-\e_3\e_3\tt)\v(t)$.  The practical complete position tracking controller then follows from applying~\eqref{eq:u1_prescription} and~\eqref{eq:u2_prescription} using~\eqref{eq:outer_claw} and~\eqref{eq:outer_claw_2}. 

A fair critique of the proposed controller and the associated guarantees is that Theorem~\ref{thrm:asymptotic_tracking} fails to provide any convergence rate that specifies when its practical tracking guarantee becomes manifest in the system's response.  For multirotor UAVs, this critique can be addressed through the use of a reference model.  Consider the following subtle interpretation of Theorem~\ref{thrm:asymptotic_stabilization}.  Referring to~\eqref{eq:stable_ball}, suppose that $V(0,\eta(0),\omega(0))=0$.  Then the ball $\calB_0 = \gamma(\bar\g r)\bar\BB^6$ is GAS and consequently forward invariant according to Theorem~\ref{thrm:asymptotic_stabilization}.  Further suppose that the initial error $e(0)$ lies within the ball $\calB_0$.  Then the practical guarantee $\|e(t)\|\leq\gamma(\bar\g r)$ is active at the initial time and remains active thereafter.  If $V(0,\eta(0),\omega(0))=0$ and $e(0)\in\calB_0$ could always be ensured, then the practical tracking guarantee would always be in effect.  The absence of a convergence rate in Theorem~\ref{thrm:asymptotic_tracking} would be moot, in this case.  In fact, $V(0,\eta(0),\omega(0))=0$ and $e(0)\in\calB_0$ can, in a sense, always be ensured.  These two conditions can be achieved by selecting $x\str(t)$ to be the output of a linear reference model whose initial conditions are selected to agree with $x(0)$, $R(0)$, and $\omega(0)$.  This reference model can then be driven by a desired outer-loop signal $x_d(t)$.  The convergence rate of the practical convergence of $x(t)$ to a neighborhood of $x_d(t)$ is then governed by the convergence rate of the reference model which may be freely designed.  

A second fair critique of the proposed controller is that the result obtained is practical, not exact, so tracking performance might be qualitatively poor.  This critique is addressed by the results of the simulation in the next section.  

\section{Simulation Example} \label{sec:Simulation}

The simulation example presented in this section depicts a multirotor UAV performing agile maneuvers\footnote{Further resources including the definition of the smooth reference position history, the code that simulates the example, and an implementation of the control law are available at \url{https://github.com/ianjwl/gimbaled-complete-tracking}.}.  The simulation is designed to demonstrate the theoretical results of Section~\ref{subsec:long_term}, illustrate the quality of tracking achievable with the \textit{practical} tracking guarantee, and demonstrate maneuvers not previously possible with state-of-the-art multirotor controllers.  Within the present context, the term agile is used to refer to maneuvering that is high acceleration and not necessarily high speed.  The translational velocity of the UAV is low enough during the simulated motions that omitting aerodynamic forces, as in~\eqref{eq:uav}, is not unreasonable.  The multirotor evolves according to~\eqref{eq:uav} with $g=9.81 \, {\rm m/s^2}$, and its inertial properties are $m = 2.09 \, {\rm kg}$ and $J=0.0297\e_1\e_1\tt + 0.0320\e_2\e_2\tt + 0.0488\e_3\e_3\tt \, {\rm kg \, m^2}$.  The parameter values selected for the outer- and inner-loop controllers are listed in Table~\ref{tab:controller_parameters}.\vspace{-0.35cm}

\begin{table}[htb!]
\begin{center}
\caption{Control parameters.}  
\label{tab:controller_parameters}
\begin{tabular}{c l c l c l}
\hline\hline
$\omega_\nu$          & $7 \; \rm rad/s$    & $\Delta_{\vaa}$         & $2 \; \rm N$                 & $\gain_\mu$    & $2 \; \rm rad/m$       \\
$\zeta_\nu$           & $1.1$               & $\Delta_{\dot n\str}$ & $2 \; {\rm s}^{-1}$ & $p_\omega$ & $1 \; \rm N \, s^2/rad^2$   \\
$r$                   & $5 \; \rm N$        & $\varepsilon$         & $10 \; \rm N$               & $\gain_\omega$ & $40 \; {\rm s}^{-1}$  \\
$U_1$                 & $2.75 \; \rm N$     & $\gain_{\vaa}$              & $2 \; {\rm s}^{-1}$ & $\gain_P$      & $2.9557 \; \rm kg/s^2$ \\
$\delta_0$            & $\pi/12 \; \rm rad$ & $\gain_\eta$              & $10 \; \rm rad/s$   & $\gain_D$      & $4.5925 \; \rm kg/s$   \\
$\delta_1$            & $\pi/6 \; \rm rad$  & $\vaa_0$                & $r\sca_{2\pi}/5$       &            &                        \\
$\varrho$             & $0.2$               & $\vaa_1$                & $r/5$        &            &                        \\
\hline
\end{tabular}
\end{center}
\end{table}\vspace{-0.35cm}

To isolate and highlight the tracking error associated with the fact that the tracking is practical, not exact, the simulation is initialized with $x(0)=x\str(0)$ and $V(0,\eta(0),\omega(0))=0$.  Furthermore, $x\str(t)$ is not processed through a reference model since the system is initialized with zero error.  The quality of the tracking performance seen in the simulation is, however, representative of the quality achievable with a reference model since the initial error condition is identical.

\begin{figure}[t!]
    \centering
    \begin{minipage}{0.48\textwidth}
        \centering
        \includegraphics[trim={0.0cm 0cm 0.0cm 0.0cm},clip,width=0.99\textwidth]{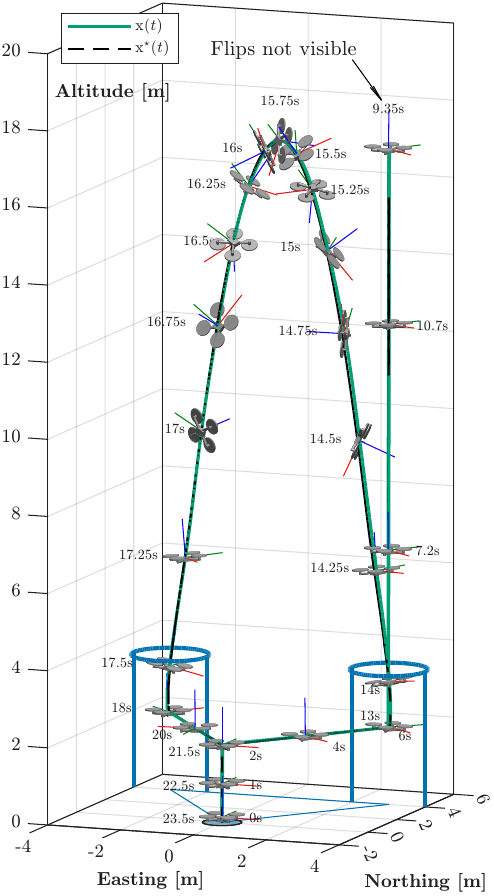}\vspace{-0.35cm}
        \caption{Simulated multirotor trajectory; see Fig.~\ref{fig:uprightness} for a visualization of the flips around $t=9.35 \, {\rm s}$.}
        \label{fig:trajectory}
    \end{minipage}\vspace{-0.5cm}
\end{figure}

\begin{figure}[t!]
    \centering
    \begin{minipage}{0.48\textwidth}
        \centering
        \includegraphics[trim={0.0cm 0cm 0.0cm 0.0cm},clip,width=0.99\textwidth]{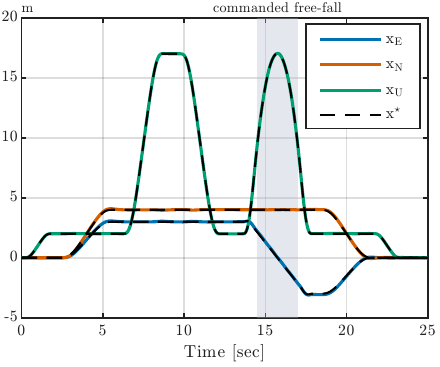}\vspace{-0.35cm}
        \caption{Position history.}
        \label{fig:position_history}
    \end{minipage}\vspace{-0.25cm}
\end{figure}

\begin{figure}[t!]
    \centering
    \begin{minipage}{0.48\textwidth}
        \centering
        \includegraphics[trim={0.0cm 0cm 0.0cm 0.0cm},clip,width=0.99\textwidth]{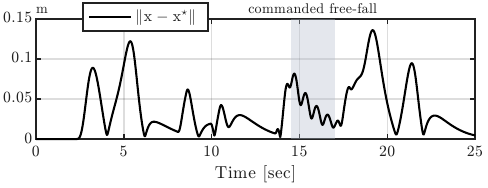}\vspace{-0.35cm}
        \caption{Position error norm history.}
        \label{fig:error}
    \end{minipage}\vspace{-0.5cm}
\end{figure}

The quality of the position tracking is illustrated in Figs.~\ref{fig:trajectory},~\ref{fig:position_history}, and~\ref{fig:error}.  Figures~\ref{fig:trajectory} and~\ref{fig:position_history} illustrate the smooth desired inertial position history $\x\str$ with dashed black curves and the position response of the multirotor with colored contours.  Figure~\ref{fig:error} depicts the norm of the error between the target and actual position histories.  The peak position error norm is approximately $15 \, {\rm cm}$ for a position history that includes agile motions on the scale of tens of meters.  With the tuning of Table~\ref{tab:controller_parameters}, the tracking quality is high and appropriate for many applications.  It is worthwhile to note that the performance can become quite sensitive for some choices of the controller parameters.  For example, performance near free fall degrades when the thrust error tolerance $r$ is decreased toward zero. The case $r=0$ corresponds to the discontinuous behavior of current state-of-the-art controllers near free fall.  

The desired position history simulates a flight through a multirotor obstacle course with several agile maneuvers that are out of reach for current state-of-the-art multirotor controllers.  Figure~\ref{fig:trajectory} includes depictions of two obstacles  ---  horizontal hoops the UAV must fly through.  Both are $2 \, {\rm m}$ in diameter and are located $3.5 \, {\rm m}$ above the ground.  There are four phases of the flight: takeoff, vertical maneuver, ballistic arc, and landing. The vertical maneuver and ballistic arc are impossible for prior state-of-the-art multirotor controllers that enjoy tracking performance guarantees. Approximations of these motions can be performed by current controllers, but following the exact commanded motions would cause these alternative controllers to encounter the free-fall singularity and fail. An expert human pilot or a learning-based controller might approximate these maneuvers with even greater accuracy, though without any formal performance guarantee.

\begin{figure}[t!]
    \centering
    \begin{minipage}{0.48\textwidth}
        \centering
        \includegraphics[trim={0.0cm 0cm 0.0cm 0.0cm},clip,width=0.99\textwidth]{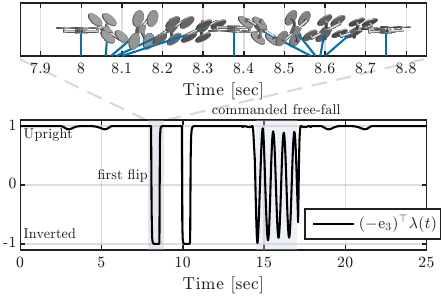}\vspace{-0.35cm}
        \caption{Uprightness of the multirotor and a visualization of one of the two flips performed during the vertical maneuver.}
        \label{fig:uprightness}
    \end{minipage}\vspace{-0.5cm}
\end{figure}

The vertical maneuver starts after the multirotor has reached the first loop obstacle.  Six seconds into the flight, the multirotor starts an aggressive $15 \, {\rm m}$ climb, followed by a $1.5 \, {\rm s}$ pause at the top of the path, and a similarly aggressive descent to the original altitude.  This vertical maneuver is agile  ---  the time allotted for the climb and descent is short enough that the motion demands a peak thrust acceleration just above $2.5g$.  The maneuver is entirely vertical, so the vehicle is commanded to fly through free fall twice: once during the climb and once during the descent.  Across each of these momentary commanded free falls, the aircraft flips to point its thrust vector in the appropriate direction.  The first of these flips is illustrated in Fig.~\ref{fig:uprightness} along with the uprightness, $(-\e_3)\tt\lambda(t)$, of the vehicle throughout the flight.  During the first flip, the vehicle is upright at $t=8 \, {\rm s}$, flips halfway over around $8.1 \, {\rm s}$, remains inverted from approximately $8.15 \, {\rm s}$ to $8.55 \, {\rm s}$, completes the flip around $8.6 \, {\rm s}$, and is again upright at $8.75 \, {\rm s}$.  The blue lines connecting each multirotor rendered in the cutout to the horizontal axis encode the time associated with the rendered attitude.

\begin{figure}[t!]
    \centering
    \begin{minipage}{0.48\textwidth}
        \centering
        \includegraphics[trim={0.0cm 0cm 0.0cm 0.0cm},clip,width=0.99\textwidth]{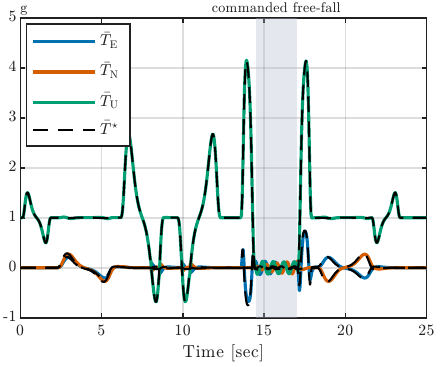}\vspace{-0.35cm}
        \caption{Normalized thrust history, $\bar T(t)=\lambda(t) u_1(t)/m$.}
        \label{fig:thrust_history}
    \end{minipage}\vspace{-0.25cm}
\end{figure}

\begin{figure}[t!]
    \centering
    \begin{minipage}{0.48\textwidth}
        \centering
        \includegraphics[trim={0.0cm 0cm 0.0cm 0.0cm},clip,width=0.99\textwidth]{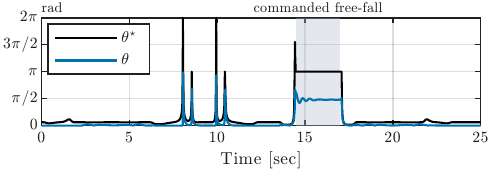}\vspace{-0.35cm}
        \caption{Comparison of $\theta(t)$ and $\theta\str(\|\vaa\str(t)\|/r)$.}
        \label{fig:angle_comparison}
    \end{minipage}\vspace{-0.25cm}
\end{figure}

Once the vertical maneuver is complete, the ballistic arc begins.  This segment of the flight begins with the multirotor applying just over $4g$ of thrust to accelerate itself through the first hoop for a ballistic arc.  For $t\in[14.5,17]\,{\rm s}$, the target position history demands that the multirotor free fall as it follows a parabolic ballistic arc.  This commanded free fall is illustrated in Figs.~\ref{fig:position_history},~\ref{fig:error},~\ref{fig:thrust_history},~\ref{fig:angle_comparison}, and~\ref{fig:uprightness} by a gray background.  Finally, the multirotor is commanded to catch itself as it flies through the second hoop.  In this simulation example, $U_1>0$ so it is impossible for the multirotor to truly free fall.  The thrust vector oscillates around zero during the commanded free fall, visible as $\bar T(t)=\lambda(t) u_1(t)/m$ in Fig.~\ref{fig:thrust_history}, as the aircraft rotates to keep the nonzero thrust from pushing it off course.  

The proposed controller provides desirable behavior when operating near free fall.  During the vertical maneuver, crossings through zero commanded thrust cause the vehicle to flip over to redirect the thrust vector.  During the commanded free fall of the ballistic arc, the thrust is reduced to the minimum allowable value and the aircraft spins as it keeps the nonzero thrust from pushing it off course.  Furthermore, boundedness of the error demonstrates the theoretical results of Section~\ref{subsec:long_term}.  To summarize, the tracking quality is desirable and the controller has been demonstrated to operate effectively near, around, and through free fall.  

\section{Conclusions} \label{sec:Conclusions}

A novel controller redesign for systems with three-dimensional gimbaled unidirectional actuation that enables practical tracking with formal guarantees was presented.  The results apply to a large class of vehicle systems.  A Lyapunov-based argument proved that the control redesign delivers practical complete tracking for the outer-loop subsystem.  In simulation, the proposed practical control design delivered desirable tracking performance and there were no obvious qualitative deficiencies relative to exact tracking.  

There are many avenues for future work.  The proposed control scheme can be augmented with adaptive modifications to the outer-loop, the inner-loop, or both to enable disturbance rejection without high gain.  It can alternatively be modified to ensure ISpS to exogenous disturbances for the outer-loop and gimbal systems.  It is also worthwhile to consider the case where the unidirectional actuation is bounded above as well as below. Robustness guarantees in the form of strong integral ISpS guarantees can likely be developed for the case of finite control authority.

\appendices

\section{Smooth Step Function}
\label{app:zeta}

The paper uses several bespoke functions defined in this appendix.  Define $\varpi:\RR\to[0,1)$ by\vspace{-0.075cm}
\begin{equation}
    \varpi(s) = \left\{ \matl{0, & s\leq0 \\ \exp(-s^{-1}), & s>0} \right.\vspace{-0.075cm}
\end{equation}
Note that $\varpi$ is $C^{\infty}$ and all of its derivatives at zero are zero.  Define $\zeta_0:\RR\to[0,1]$ by\vspace{-0.075cm}
\begin{equation}
    \zeta_0(s) = \frac{\varpi(s)}{\varpi(s)+\varpi(1-s)}\vspace{-0.075cm}
\end{equation}
Note that $\zeta_0$ is a $C^{\infty}$  step function with $\zeta_0(s)=0$ for all $s\leq0$ and $\zeta_0(s)=1$ for all $s\geq1$.  Define $\zeta:\RR\times\RR\times\RR\to[0,1]$ by\vspace{-0.075cm}
\begin{equation}
    \zeta(s;s_0,s_1) = \zeta_0\bigl(\frac{s-s_0}{s_1-s_0}\bigr)\vspace{-0.075cm}
\end{equation}
The function $\zeta(s;s_0,s_1)$ is a $C^{\infty}$ step function.  Its argument is $s$ and $s_0,s_1$ are parameters. Respectively, $s_0$ and $s_1$ are the values of $s$ where the value of $\zeta$ becomes $0$ and $1$.  If $s_0=s_1$, then $\zeta$ is not defined.  If $s_0<s_1$, then the value of $\zeta$ is zero to the left of the interval $(s_0,s_1)$ and is one to the right of it.  If $s_1<s_0$, then the value of $\zeta$ is one to the left of the interval $(s_1,s_0)$ and is zero to the right of it. An identity that can be used to simplify the notation of equations like~\eqref{eq:defn_of_theta_star} is\vspace{-0.075cm}
\begin{equation}
    1-\zeta(s;s_0,s_1) = \zeta(s;s_1,s_0)\vspace{-0.075cm}
\end{equation}
For clarity, however, we avoid using this identity and instead always write $\zeta(s;s_0,s_1)$ with $s_0 < s_1$.

Finally, define $Z:\RR\times\RR\times\RR\to\RR_{\geq0}$ by\vspace{-0.075cm}
\begin{equation}
    Z(s;s_0,s_1) = \dintt{s_0}{s}{\zeta(\sigma;s_0,s_1)}{\sigma}\vspace{-0.075cm}
\end{equation}
The function $Z(s;s_0,s_1)$ is a smooth ramp function based on $\zeta(s;s_0,s_1)$.  For $s_0<s_1$, $Z(s;s_0,s_1)=0$ for $s\leq s_0$ and $Z(s;s_0,s_1)=s - \tfrac{1}{2}(s_0+s_1)$ for $s\geq s_1$.   For $s_0=s_1$, we define $Z(s;s_0,s_1)=\max\{0,s - \tfrac{1}{2}(s_0+s_1)\}$.

\section{Deferred Proofs}\label{app:proofs}

\subsection{Proof of Lemma~\ref{lemma:angle_bound_on_dist}}
\label{appsec:angle_bound_on_dist}
\begin{proof}
    Let $d:\RR_{\geq0}\times[0,\pi]\to\RR_{\geq0}$ be defined by\vspace{-0.075cm}
    \begin{equation} \label{eq:defn_of_d}
        d(\sca,\phi) = r\max\Bigl\{0,\bigl\|\kappa(\sca,\phi)\bsm{\cos\phi\\\sin\phi}-\sca\e_1\bigr\|-1\Bigr\}.\vspace{-0.075cm}
    \end{equation}
    Along trajectories, $\|\vaa(t)\|_{\setA(t)} = d(\|\vaa\str(t)\|/r,\theta(t))$ whenever $\theta(t)$ is defined. Recall that $\vaa\str(t)=0\Rightarrow\|\vaa(t)\|_{\setA(t)}=0$ has been previously established.  To prove the remaining claim of Lemma~\ref{lemma:angle_bound_on_dist}, we show that\vspace{-0.075cm}
    \begin{equation}
        d(\sca,\phi) \leq \ell_{\vaa} r\, \sca\max\{0,\phi-\theta\str(\sca)\}
        \label{eq:intermediate_angle_bound}\vspace{-0.075cm}
    \end{equation}
    for all $(\sca,\phi)\in\RR_{\geq0}\times[0,\pi]$, for some $\ell_{\vaa}>0$.  

    Note that the definitions~\eqref{eq:intersection_length},~\eqref{eq:defn_of_kappa_circ}, and~\eqref{eq:defn_of_kappa} ensure that\vspace{-0.075cm}
    \begin{equation}
        \kappa(\sca,\phi) \leq \kappa_\circ(\sca) \leq l\str_+(\sca) \leq \sca+1\vspace{-0.075cm}
    \end{equation}
    for all $(\sca,\phi)\in\RR_{\geq0}\times[0,\pi]$.  The vector $\kappa(\sca,\phi)\bsm{\cos\phi\\\sin\phi}$ therefore lies within a radius $\sca+1$ of the origin.  Hence,\vspace{-0.075cm}
    \begin{equation}
        \bigl\|\kappa(\sca,\phi)\bsm{\cos\phi\\\sin\phi}\!-\!\sca\e_1\bigr\| \leq \!\!\sup_{c\,\in (\sca+1)\bar\BB^2}\!\! \|c - \sca\e_1\| = 2\sca + 1.\vspace{-0.075cm}
    \end{equation}
    Applying this bound to the normed vector in~\eqref{eq:defn_of_d} reveals that
    \begin{equation}
        d(\sca,\phi) \leq r\max\bigl\{0,(2\sca+1)-1\bigr\} = 2r\sca.\vspace{-0.075cm}
        \label{eq:simple_distance_bound}
    \end{equation}
    This overestimate of $d(\sca,\phi)$ is later used to design the constant $\ell_{\vaa}$ that appears in~\eqref{eq:intermediate_angle_bound}.

    Next, note that $d(\sca,\phi)=0$ whenever $\phi\leq\theta\str(\sca)$. Suppose $\sca$ is small enough that $\theta\str(\sca)\geq\pi$.  Then $d(\sca,\phi)=0$ for all $(\sca,\phi)\in\RR_{\geq0}\times[0,\pi]$, so~\eqref{eq:intermediate_angle_bound} holds.  Suppose that $\sca$ is large enough that $\theta\str(\sca)<\pi$.  Then $d(\sca,\phi)=0$ whenever $\phi\in[0,\theta\str(\sca)]$.  Consider the interval $(\theta\str(\sca),\min\{\theta\str(\sca)+\delta_0,\pi\})$ and the derivative of $d(\sca,\phi)$ with respect to $\phi$.  For $\|\kappa(\sca,\phi)\bsm{\cos\phi\\\sin\phi}-\sca\e_1\bigr\|<1$, the derivative is zero.  For $\|\kappa(\sca,\phi)\bsm{\cos\phi\\\sin\phi}-\sca\e_1\bigr\|>1$, the derivative is\vspace{-0.075cm}
    \begin{equation}
        \tfrac{\partial}{\partial\phi} d(\sca,\phi) = r\Bigl(\tfrac{\kappa(\sca,\phi) \, \partial_\phi \kappa(\sca,\phi) - \sca\partial_\phi \kappa(\sca,\phi) \cos\phi + \sca \kappa(\sca,\phi) \sin\phi}{\|\kappa(\sca,\phi)\bsm{\cos\phi\\\sin\phi}-\sca\e_1\bigr\|}\Bigr).
        \label{eq:derivative_of_d_v1}\vspace{-0.075cm}
    \end{equation}
    From the definition of $\kappa$ in~\eqref{eq:defn_of_kappa}, we have $\kappa(\sca,\phi)=\kappa_\circ(\sca)$ and $\partial_\phi \kappa(\sca,\phi)=0$ for $\phi\in(\theta\str(\sca),\theta\str(\sca)+\delta_0)$.  With this fact, the derivative in~\eqref{eq:derivative_of_d_v1} can be written as\vspace{-0.075cm}
    \begin{equation}
        \tfrac{\partial}{\partial\phi} d(\sca,\phi) = \frac{r\sca \, \kappa_\circ(\sca) \sin\phi}{\|\kappa(\sca,\phi)\bsm{\cos\phi\\\sin\phi}-\sca\e_1\bigr\|}.
        \label{eq:derivative_of_d_v2}\vspace{-0.075cm}
    \end{equation}
    From here, the argument presented in Appendix~\ref{appsec:derivative_bound} ensures\vspace{-0.075cm}
    \begin{equation}
        \tfrac{\partial}{\partial\phi} d(\sca,\phi) = \frac{r\sca \, \kappa_\circ(\sca) \sin\phi}{\sqrt{\kappa_\circ(\sca)^2 - 2 \,\sca\, \kappa_\circ(\sca) \cos\phi + \sca^2}} \leq r\sca
        \label{eq:bound_on_derivative_of_d}\vspace{-0.075cm}
    \end{equation}
    for $\phi\in(\theta\str(\sca),\theta\str(\sca)+\delta_0)$.  
    
    Exclusively considering the restricted interval $[0,\min\{\theta\str(\sca)+\delta_0,\pi\})$,~\eqref{eq:bound_on_derivative_of_d} directly guarantees\vspace{-0.075cm}
    \begin{equation}
        d(\sca,\phi) \leq r\sca \max\{0,\phi-\theta\str(\sca)\}
        \label{eq:limited_angle_bound}\vspace{-0.075cm}
    \end{equation}
    on this interval due to continuity and the fact that $d(\sca,\phi)=0$ for all $\phi\in[0,\theta\str(\sca)]$.  To ensure the bound~\eqref{eq:intermediate_angle_bound} holds over the entire interval $[0,\pi]$ (rather than just over $[0,\min\{\theta\str(\sca)+\delta_0,\pi\})$ with $\ell_{\vaa}=1$), prescribe\vspace{-0.075cm}
    \begin{equation}
        \ell_{\vaa} = 2\max\{\delta_0^{-1},1\}.\vspace{-0.075cm}
    \end{equation}
    This choice ensures that $\ell_{\vaa} r\sca \max\{0,\phi-\theta\str(\sca)\}$ reaches or exceeds the upper bound $2r\sca$ on $d(\sca,\phi)$ seen in~\eqref{eq:simple_distance_bound} by the end of the interval $[0,\min\{\pi,\theta\str(\sca)+\delta_0\})$.  If the interval does not include the angle $\theta\str(\sca)+\delta_0$ because this sum happens to be greater than $\pi$, then the bound $\ell_{\vaa} \, r\sca \max\{0,\phi-\theta\str(\sca)\}$ may not reach $2r\sca$ by $\phi=\pi$, but it will still overestimate $d(\sca,\phi)$ for all $(\sca,\phi)\in\RR_{\geq0}\times[0,\pi]$.
\end{proof}

\subsection{Derivative Bound in the Proof of Lemma~\ref{lemma:angle_bound_on_dist}}
\label{appsec:derivative_bound}

\begin{proposition}
    The bound\vspace{-0.075cm}
    \begin{equation}
        \mathfrak{F}(\sca,\phi) \coloneqq \frac{\sca \, \kappa_\circ(\sca) \sin\phi}{\bigl\|\kappa(\sca,\phi)\bbm{\cos\phi & \sin\phi}\tt-\sca\e_1\bigr\|} \leq \sca \label{eq:derivative_bound}\vspace{-0.075cm}
    \end{equation}
    holds for all $\sca>0$, whenever $\phi\in(\theta\str(\sca),\theta\str(\sca)+\delta_0)$.
\end{proposition}

\begin{proof}
    Directly evaluating the norm in the denominator of $\mathfrak{F}(\sca,\phi)$ yields\vspace{-0.075cm}
    \begin{equation}
        \bigl\|\kappa(\sca,\phi)\bsm{\cos\phi \\ \sin\phi}-\sca\e_1\bigr\| = \sqrt{\kappa(\sca,\phi)^2 - 2 \sca \kappa(\sca,\phi) \cos\phi + \sca^2}.\vspace{-0.075cm}
    \end{equation}
    Recall that $\kappa(\sca,\phi)=\kappa_\circ(\sca)$ for all $\phi\in[0,\theta\str(\sca)+\delta_0]$ according to the definition of $\kappa(\sca,\phi)$ in~\eqref{eq:defn_of_kappa}.  On the interval of interest $\phi\in(\theta\str(\sca),\theta\str(\sca)+\delta_0)$, therefore,\vspace{-0.075cm}
    \begin{equation}
        \mathfrak{F}(\sca,\phi) = \frac{\sca\, \kappa_\circ(\sca) \sin\phi}{\sqrt{\kappa_\circ(\sca)^2 - 2 \sca \kappa_\circ(\sca) \cos\phi + \sca^2}}\vspace{-0.075cm}
    \end{equation}
    From this form of $\mathfrak{F}(\sca,\phi)$, it follows that\vspace{-0.075cm}  
    \begin{align*}
        &
        \mathfrak{F}(\sca,\phi) = \frac{\sca \kappa_\circ(\sca) \sin\phi}{\sqrt{\kappa_\circ(\sca)^2 - 2 \sca \kappa_\circ(\sca) \cos\phi + \sca^2}} \leq \sca
        \\&\Leftrightarrow\quad
        \kappa_\circ(\sca) \sin\phi \leq \sqrt{\kappa_\circ(\sca)^2 - 2 \sca \kappa_\circ(\sca) \cos\phi + \sca^2}
        \\&\Leftrightarrow\quad
        \kappa_\circ(\sca)^2 \sin^2\phi \leq \kappa_\circ(\sca)^2 - 2 \sca \kappa_\circ(\sca) \cos\phi + \sca^2
        \\&\Leftrightarrow\quad
        \kappa_\circ(\sca)^2(1-\cos^2\phi) \leq \kappa_\circ(\sca)^2 - 2 \sca \kappa_\circ(\sca) \cos\phi + \sca^2
        \\&\Leftrightarrow\quad
        0 \leq (\kappa_\circ(\sca)\cos\phi)^2 - 2 (\kappa_\circ(\sca) \cos\phi)\sca + \sca^2
        \\&\Leftrightarrow\quad
        0 \leq (\kappa_\circ(\sca)\cos\phi - \sca)^2 \quad \text{which is a fact,}\vspace{-0.075cm}
    \end{align*}
    where we have used the conditions $\sca>0$ and $\phi > \theta\str(\sca)$ from the proposition and the facts that $\theta\str(\sca)>0$ and $\kappa_\circ(\sca)\geq0$ from their definitions.
\end{proof}

\section*{References}
\bibliographystyle{IEEEtran}
\bibliography{references}

\begin{thebibliography}{10}
\providecommand{\url}[1]{#1}
\csname url@samestyle\endcsname
\providecommand{\newblock}{\relax}
\providecommand{\bibinfo}[2]{#2}
\providecommand{\BIBentrySTDinterwordspacing}{\spaceskip=0pt\relax}
\providecommand{\BIBentryALTinterwordstretchfactor}{4}
\providecommand{\BIBentryALTinterwordspacing}{\spaceskip=\fontdimen2\font plus
\BIBentryALTinterwordstretchfactor\fontdimen3\font minus \fontdimen4\font\relax}
\providecommand{\BIBforeignlanguage}[2]{{%
\expandafter\ifx\csname l@#1\endcsname\relax
\typeout{** WARNING: IEEEtran.bst: No hyphenation pattern has been}%
\typeout{** loaded for the language `#1'. Using the pattern for}%
\typeout{** the default language instead.}%
\else
\language=\csname l@#1\endcsname
\fi
#2}}
\providecommand{\BIBdecl}{\relax}
\BIBdecl

\bibitem{willebeek_lemair_practical_2026}
\BIBentryALTinterwordspacing
I.~J. Willebeek-LeMair and C.~A. Woolsey, ``Practical {Complete} {Tracking} {With} {Pivoted} {Unidirectional} {Actuation},'' Sep. 2026, [eess.SY]. [Online]. Available: \url{http://arxiv.org/abs/2604.05979}
\BIBentrySTDinterwordspacing

\bibitem{minh-duc_hua_control_2009}
M.-D. Hua, T.~Hamel, P.~Morin, and C.~Samson, ``A {Control} {Approach} for {Thrust}-{Propelled} {Underactuated} {Vehicles} and its {Application} to {VTOL} {Drones},'' \emph{IEEE Transactions on Automatic Control}, vol.~54, no.~8, pp. 1837--1853, 2009.

\bibitem{lee_geometric_2010}
T.~Lee, M.~Leok, and N.~H. McClamroch, ``Geometric tracking control of a quadrotor {UAV} on {SE}(3),'' in \emph{49th {IEEE} {Conference} on {Decision} and {Control} ({CDC})}.\hskip 1em plus 0.5em minus 0.4em\relax Atlanta, GA, USA: IEEE, 2010, pp. 5420--5425.

\bibitem{lee_control_2011}
------, ``Control of {Complex} {Maneuvers} for a {Quadrotor} {UAV} using {Geometric} {Methods} on {SE}(3),'' 2011, [math].

\bibitem{lee_geometric_2011}
------, ``Geometric {Tracking} {Control} of a {Quadrotor} {UAV} for {Extreme} {Maneuverability},'' \emph{IFAC Proceedings Volumes}, vol.~44, 2011.

\bibitem{mellinger_minimum_2011}
D.~Mellinger and V.~Kumar, ``Minimum snap trajectory generation and control for quadrotors,'' in \emph{2011 {IEEE} {International} {Conference} on {Robotics} and {Automation}}, 2011, pp. 2520--2525.

\bibitem{lee_nonlinear_2013}
T.~Lee, M.~Leok, and N.~H. McClamroch, ``Nonlinear {Robust} {Tracking} {Control} of a {Quadrotor} {UAV} on {SE}(3),'' \emph{Asian Journal of Control}, vol.~15, pp. 391--408, 2013.

\bibitem{gamagedara_geometric_2019}
K.~Gamagedara, M.~Bisheban, E.~Kaufman, and T.~Lee, ``Geometric {Controls} of a {Quadrotor} {UAV} with {Decoupled} {Yaw} {Control},'' in \emph{2019 {American} {Control} {Conference} ({ACC})}, 2019.

\bibitem{flores_robust_2025}
G.~Flores, A.~M. Boker, M.~Al~Janaideh, and M.~W. Spong, ``Robust {Output} {Feedback} {Control} with {Predefined} {State} {Boundaries} for {Multi}-{Rotor} {Systems},'' in \emph{2025 {IEEE} 64th {Conference} on {Decision} and {Control} ({CDC})}, 2025, pp. 831--836.

\bibitem{willebeek_lemair_input_state_2025}
I.~Willebeek-LeMair, S.~B. Widman, and C.~A. Woolsey, ``Input-to-{State} {Stable} {Energy}-{Based} {Position} {Tracking} {Control} for {Atmospheric} {Flight} {Vehicles},'' in \emph{{AIAA} {SCITECH} 2025 {Forum}}.\hskip 1em plus 0.5em minus 0.4em\relax American Institute of Aeronautics and Astronautics, 2025.

\bibitem{willebeek_lemair_robust_2025}
I.~J. Willebeek-LeMair and C.~A. Woolsey, ``Robust {Port}-{Hamiltonian} {Output}-{Tracking} {Control} of {Cascaded} {Systems},'' in \emph{2025 {IEEE} 64th {Conference} on {Decision} and {Control} ({CDC})}, 2025.

\bibitem{casau_robust_2015}
P.~Casau, R.~G. Sanfelice, R.~Cunha, D.~Cabecinhas, and C.~Silvestre, ``Robust global trajectory tracking for a class of underactuated vehicles,'' \emph{Automatica}, vol.~58, pp. 90--98, 2015.

\bibitem{naldi_robust_2017}
R.~Naldi, M.~Furci, R.~G. Sanfelice, and L.~Marconi, ``Robust {Global} {Trajectory} {Tracking} for {Underactuated} {VTOL} {Aerial} {Vehicles} {Using} {Inner}-{Outer} {Loop} {Control} {Paradigms},'' \emph{IEEE Transactions on Automatic Control}, vol.~62, pp. 97--112, 2017.

\bibitem{invernizzi_trajectory_2018}
D.~Invernizzi and M.~Lovera, ``Trajectory tracking control of thrust-vectoring {UAVs},'' \emph{Automatica}, vol.~95, pp. 180--186, 2018.

\bibitem{casau_robust_2019}
P.~Casau, C.~G. Mayhew, R.~G. Sanfelice, and C.~Silvestre, ``Robust global exponential stabilization on the n-dimensional sphere with applications to trajectory tracking for quadrotors,'' \emph{Automatica}, vol. 110, 2019.

\bibitem{invernizzi_dynamic_2020}
D.~Invernizzi, M.~Lovera, and L.~Zaccarian, ``Dynamic {Attitude} {Planning} for {Trajectory} {Tracking} in {Thrust}-{Vectoring} {UAVs},'' \emph{IEEE Transactions on Automatic Control}, vol.~65, no.~1, pp. 453--460, 2020.

\bibitem{martins_global_2021}
L.~Martins, C.~Cardeira, and P.~Oliveira, ``Global trajectory tracking for quadrotors: {An} {MRP}-based hybrid backstepping strategy,'' in \emph{2021 60th {IEEE} {Conference} on {Decision} and {Control} ({CDC})}, 2021.

\bibitem{martins_robust_2023}
------, ``Robust {Global} {Exponential} {Trajectory} {Tracking} for {Quadrotors}: {An} {MRP}-{Based} {Hybrid} {Approach},'' in \emph{2023 62nd {IEEE} {Conference} on {Decision} and {Control} ({CDC})}, 2023, pp. 5249--5254.

\bibitem{martins_global_2024}
------, ``Global trajectory tracking for quadrotors: {An} {MRP}-based hybrid strategy with input saturation,'' \emph{Automatica}, vol. 162, 2024.

\bibitem{leomanni_almost_2024}
M.~Leomanni, A.~Dionigi, F.~Ferrante, P.~Valigi, and G.~Costante, ``Almost {Global} {Trajectory} {Tracking} for {Quadrotors} {Using} {Thrust} {Direction} {Control} on {S2},'' in \emph{2024 {IEEE} 63rd {Conference} on {Decision} and {Control} ({CDC})}, 2024, pp. 7516--7521.

\bibitem{ballaben_towards_2025}
R.~Ballaben, A.~Astolfi, P.~Braun, and L.~Zaccarian, ``Towards global stabilization of a hovercraft model using hybrid systems and discontinuous feedback laws,'' in \emph{2025 {IEEE} 64th {Conference} on {Decision} and {Control} ({CDC})}, 2025, pp. 6579--6584.

\bibitem{bahati_control_2025}
G.~Bahati, R.~K. Cosner, M.~H. Cohen, R.~M. Bena, and A.~D. Ames, ``Control {Barrier} {Function} {Synthesis} for {Nonlinear} {Systems} with {Dual} {Relative} {Degree},'' in \emph{2025 {IEEE} 64th {Conference} on {Decision} and {Control} ({CDC})}, 2025, pp. 8208--8215.

\bibitem{lin_various_1995}
Y.~Lin, E.~Sontag, and Y.~Wang, ``Various results concerning set input-to-state stability,'' in \emph{Proceedings of 1995 34th {IEEE} {Conference} on {Decision} and {Control}}, vol.~2, 1995, pp. 1330--1335.

\bibitem{kim_global_2026}
\BIBentryALTinterwordspacing
K.~H. Kim, V.~Todorovski, and M.~Krstic, ``Global {Exponential} {Stabilization} of a {3D} {Nonholonomic} {Vehicle} in {Spherical} {Coordinates},'' Sep. 2026, [math.OC]. [Online]. Available: \url{http://arxiv.org/abs/2609.24192}
\BIBentrySTDinterwordspacing

\bibitem{khalil_nonlinear_2002}
H.~Khalil, \emph{Nonlinear {Systems}}, 3rd~ed.\hskip 1em plus 0.5em minus 0.4em\relax Prentice Hall, 2002.

\bibitem{liberzon_switching_2003}
D.~Liberzon, \emph{Switching in {Systems} and {Control}}, ser. Systems \& {Control}: {Foundations} \& {Applications}, T.~Başar, Ed.\hskip 1em plus 0.5em minus 0.4em\relax Birkhäuser, 2003.

\bibitem{goebel_hybrid_2009}
R.~Goebel, R.~G. Sanfelice, and A.~R. Teel, ``Hybrid dynamical systems,'' \emph{IEEE Control Systems Magazine}, vol.~29, 2009.

\bibitem{chaillet_uniform_2008}
A.~Chaillet and A.~Loría, ``Uniform semiglobal practical asymptotic stability for non-autonomous cascaded systems and applications,'' \emph{Automatica}, vol.~44, no.~2, pp. 337--347, 2008.

\end{thebibliography}

\begin{IEEEbiography}[{\includegraphics[width=1in,height=1.25in,clip,keepaspectratio]{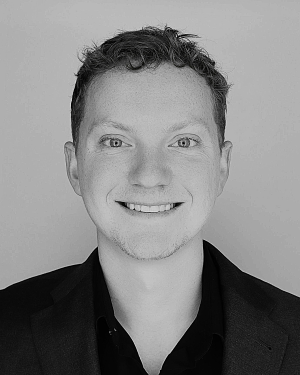}}]{Ian Willebeek-LeMair} (Student Member, IEEE) received the B.S. degree in aerospace and ocean engineering from Virginia Tech, Blacksburg, VA, USA, in 2021.  

He is currently a Ph.D. candidate with the Kevin T. Crofton Department of Aerospace and Ocean Engineering, Virginia Tech, Blacksburg, VA, USA. Previously, he was a flight controls engineer with NAVAIR from 2021 to 2023.  His research focuses on agile and robust nonlinear control of air, ocean, and space vehicles.

Mr. Willebeek-LeMair is a student member of the Control Systems Society and the AIAA.  His 2025 paper ``Robust Port-Hamiltonian Output-Tracking Control of Cascaded Systems'' received the IEEE CSS Manufacturing Automation and Robotics Control Technical Committee's outstanding paper award for the 2025 Conference on Decision and Control.  
\end{IEEEbiography}\vspace{-0.05cm}

\begin{IEEEbiography}[{\includegraphics[width=1in,height=1.25in,clip,keepaspectratio]{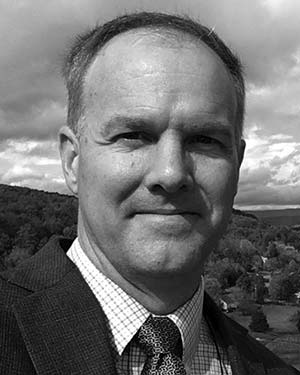}}]{Craig Woolsey} (Senior Member, IEEE) received the B.M.E. degree in Bachelor of Mechanical Engineering from Georgia Tech, Atlanta, GA, USA, in 1995, and the Ph.D. degree in mechanical and aerospace engineering from Princeton University, Princeton, NJ, USA, in 2001.

He is currently a Professor with the Kevin T.\ Crofton Department of Aerospace and Ocean Engineering, Virginia Tech, Blacksburg, VA, USA. His research and teaching focus on modeling and control of ocean and air vehicles.

Dr. Woolsey participates in the Control Systems Society, the Oceanic Engineering Society, and the AIAA. He is also active on the IFAC Technical Committee on Marine Systems and the IEEE CSS Manufacturing Automation and Robotics Control Technical Committee.
\end{IEEEbiography}

\end{document}